\documentclass[12pt,letterpaper]{article}

\usepackage{geometry}
\usepackage{amsmath}
\usepackage{graphicx}
\usepackage{enumerate}
\usepackage{natbib}
\usepackage{url} 
\usepackage{amssymb}
\usepackage{dsfont} 

\usepackage{xr}
\usepackage{amsthm} 
\theoremstyle{remark}  
\newtheorem{theorem}{Theorem} 
\theoremstyle{remark}  
\newtheorem{remark}{Remark} 
\usepackage{arydshln} 
\usepackage{caption} 
\usepackage[table]{xcolor}
\definecolor{light-gray}{gray}{0.8}

\pdfoutput=1 

\usepackage[utf8]{inputenc}
\usepackage{cite}
\usepackage{nameref,hyperref}
\usepackage[right]{lineno}
\usepackage{microtype}
\DisableLigatures[f]{encoding = *, family = * }

\usepackage{changepage}
\usepackage[aboveskip=1pt,labelfont=bf,labelsep=period,singlelinecheck=off]{caption}
\makeatletter
\renewcommand{\@biblabel}[1]{\quad#1.}
\makeatother

\usepackage{lastpage,fancyhdr,graphicx}
\usepackage{epstopdf}
\usepackage{natbib}
\usepackage{amsmath}
\fancyheadoffset[L]{2.25in}
\fancyfootoffset[L]{2.25in}

\usepackage{color}

\definecolor{Gray}{gray}{.25}

\usepackage{graphicx}

\usepackage{wrapfig}
\usepackage[pscoord]{eso-pic}
\usepackage[fulladjust]{marginnote}
\reversemarginpar

\usepackage{algorithm} 
\usepackage[noend]{algpseudocode}
\usepackage{enumitem}
\makeatletter
\def\BState{\State\hskip-\ALG@thistlm}
\makeatother

\begin{document}
\vspace*{0.35in}
\begin{flushleft}
{\Large
\textbf\newline{\center{The Importance of Being a Band: Finite-Sample Exact Distribution-Free Prediction Sets for Functional Data}}
}
\newline
\\
Jacopo Diquigiovanni\textsuperscript{1,*},
  Matteo Fontana\textsuperscript{2,3},
  Simone Vantini\textsuperscript{2}
\\
\bigskip
\bf{1} Department of Statistical Sciences, University of Padova, Italy
\\
\bf{2} MOX - Department of Mathematics, Politecnico di Milano, Italy
\\
\bf{3} now at Joint Research Centre - European Commission, Ispra (VA), Italy
\\
\bigskip
* jacopo.diquigiovanni@phd.unipd.it

\end{flushleft}

\section*{Abstract}
Functional Data Analysis represents a field of growing interest in statistics. Despite several studies have been proposed leading to fundamental results, the problem of obtaining valid and efficient prediction sets has not been thoroughly covered.
Indeed, the great majority of methods currently in the literature rely on strong distributional assumptions (e.g, Gaussianity), dimension reduction techniques and/or asymptotic arguments. In this work, we propose a new nonparametric approach in the field of Conformal Prediction based on a new family of nonconformity measures inducing conformal predictors able to create closed-form finite-sample valid or exact prediction sets under very minimal distributional assumptions. In addition, our proposal ensures that the prediction sets obtained are bands, an essential feature in the functional setting that allows the visualization and interpretation of such sets. The procedure is also fast, scalable,  does not rely on functional dimension reduction techniques and allows the user to select different nonconformity measures depending on the problem at hand always obtaining valid bands. Within this family of measures, we propose also a specific measure leading to prediction bands asymptotically no less efficient than those with constant width.

\noindent%

{\it Keywords:}  Conformal Prediction; Distribution-free prediction set; Exact prediction set; Functional data;  Prediction band; Valid prediction set

\section{Introduction}
\label{sec:introduction}

One of the main roles of statistics in our new, data-rich world is to provide scientists, business people and policy makers with tools able to deal with an increasing amount of data, of increasing complexity. Automated sensor arrays and measuring systems now provide huge quantities of high-frequency and high-dimensional data about all sorts of social or physical phenomena. 

Among the most popular toolboxes that have the capacity to deal with this kind of complex data one can find Functional Data Analysis \citep[FDA,][]{ramsay_functional_2005}. FDA is an ebullient field of statistics which aim is to develop theory and methods to deal with data sets made of functions defined over a domain, either uni- or multidimensional, and usually characterized by some degree of smoothness. 
In the following, we will indicate with $\mathcal{Y(T)}$ the family of functions $y: \mathcal{T} \rightarrow \mathbb{R}$ belonging to $L^{\infty}(\mathcal{T})$ with $\mathcal{T}$ closed and bounded subset of $\mathbb{R}^d$, $d \in \mathbb{N}_{>0}$, and with $y_1,\dots,y_n$ possible realizations of $n$ i.i.d. random functions $Y_1,\dots,Y_n \sim P$ taking values in  $\mathcal{Y(T)}$. Without loss of generality, hereafter we will consider $d=1$ since it is the most common practical case.
Despite being born in relatively recent times \citep{ramsay_when_1982}, a plethora of standard multivariate tools have ported to the functional realm: among others Functional Principal Component Analysis \citep[Chapter 10]{ramsay_functional_2005}, Functional Linear Regression  \citep[Chapter 12]{ramsay_functional_2005} and Functional Boxplots \citep{sun_functional_2011}. 

A problem that, perhaps surprisingly, has not been covered in a satisfactory way in the FDA literature is the issue of uncertainty quantification in prediction and forecasting. 
In a more formal way, the interest is in the creation of prediction sets, namely subsets of $\mathcal{Y(T)}$ that include a new function $Y_{n+1}$ (i.i.d to $Y_1,\dots,Y_n$) with a certain nominal confidence level $1-\alpha$. 
In particular, the aim is to obtain either exact - i.e. ensuring a coverage equal to the nominal confidence level - or at least valid - i.e. ensuring a coverage no less than the nominal confidence level - prediction sets. 
Recent works in FDA provide novel insights into this very meaningful applied and theoretical issue. These attempts can be broadly classified in two classes: a first one, composed of works based mainly on parametric bootstrapping techniques \citep[e.g.,][]{degras_simultaneous_2011, cao_simultaneous_2012}, and a second one, where a dimensionality reduction technique is applied to render the naturally infinite-dimensional problem more tractable by projecting it on a finite dimensional functional basis \citep[e.g.,][]{hyndman_robust_2007,antoniadis_prediction_2016}. These approaches carry some shortcomings: the first group of techniques is computationally intensive, thus requiring long calculation times, while the second ones rely on the approximations introduced by basis projection. Both of them, in any case, either rely on not easily provable distributional assumptions and/or on asymptotic results.

The framework of this manuscript is Conformal Prediction \citep{vovk2005algorithmic,shafer_tutorial_2008}, a novel method of forecasting firstly developed in the Machine Learning community as a way to define prediction intervals for Support Vector Machines \citep{gammerman_learning_1998}. The interested reader can find a recent review in \citet{zeni_conformal_2020}. In univariate setting, Conformal Prediction is able to generate distribution-free, valid prediction intervals and it has also been used as a data exploration tool for Functional Data \citep{lei2015conformal}, via the use of a truncated basis approach.

In this article, we build on top of the literature about set prediction for functional data and Conformal Prediction, by introducing several theoretical and methodological innovations.
\begin{enumerate}
\item 
In Section \ref{sec:importance_rectangular} we show the importance in interpretative terms of obtaining functional prediction sets having a specific shape (i.e. prediction bands) through a motivating example.  
\item 
In Section \ref{sec:conformal_approach} functional prediction sets are formally defined and the Semi-Off-Line Inductive Conformal framework, also known simply as Split Conformal, is introduced. Specifically, we contribute in two ways to the Conformal Prediction literature: via enriching the results about the validity of split conformal prediction sets by making the exact probability reached by them explicit (Theorem \ref{th_conservativeness_up_low_bound}) and we provide what is to the best of our knowledge the first formal proof of the exactness of smoothed split conformal prediction sets (Appendix A.1). 
\item 
In Section \ref{sec:supremum_metric} we propose a nonconformity measure inducing a conformal predictor able to create closed-form finite-sample either valid or exact prediction bands of constant amplitude, under minimal distributional assumptions. The procedure is fast, scalable and does not rely on widespread functional dimension reduction techniques. 
\item 
In Section \ref{sec:modulation} we propose a family of nonconformity measures (to which the nonconformity measure introduced in Section \ref{sec:supremum_metric} belongs) indexed by modulation function $s_{\mathcal{I}_1}$ that allows for prediction bands with non-constant width, but able to keep all the aforementioned appealing properties. As a consequence, prediction bands induced by the nonconformity measures belonging to this family can be compared on the basis of  features other than validity, such as efficiency  (i.e. the size). 
\item 
In Section \ref{sec:modulation} we focus on a specific nonconformity measure belonging to this family which leads to valid prediction bands asymptotically no less efficient than those obtained by not modulating (Theorem \ref{th:shared_convergence}, Theorem \ref{th:better_than_not_modul}).
\end{enumerate}
Finally, in Section \ref{sec:simulation_study} we propose a simulation study to compare our method with four alternatives, and in Section \ref{sec:application} we apply our approach to the Berkeley Growth Study data set \citep{tuddenham1954physical}. Section \ref{sec:discussion} provides an overview of the main results.

\section{The Importance of Being a Band}
\label{sec:importance_rectangular}
Set prediction is of key importance in the statistical community.
Specifically, three main features characterize a prediction set: shape, coverage, and size. We start by tackling, in this section, the first issue, while the last two are explored in Section \ref{sec:conformal_prediction_bands}.
In the classical multivariate statistical setting, elliptic regions have been and are still considered as the standard shapes for prediction sets. Differently, in the functional context many authors \citep{lopez2009concept,lei2015conformal} note how the focus should be on a particular type of prediction set, commonly known as \textit{prediction band}. Formally, a band is defined as
\begin{equation*}
\left\{y \in \mathcal{Y(T)}: y(t) \in B_n(t), \quad \forall t \in \mathcal{T} \right\},
\end{equation*}
with $B_n(t)  \subseteq \mathbb{R}$ interval for each $t \in \mathcal{T}$ \citep{lopez2009concept, degras2017simultaneous}. 
The focus on this type of sets, that can be defined as the Cartesian product of the (infinitely many) intervals $\left\{B_n(t) : t \in \mathcal{T} \right\}$, comes from the fact that -- differently from a generic region of $\mathcal{Y(T)}$ -- such a shape can be easily visualized on a plot (i.e., it is a band, in parallel coordinates, as noted by \citet{lopez2009concept}) and thus interpreted with respect to the domain $\mathcal{T}$. 
\begin{figure}
\begin{center}
\includegraphics[scale=0.7]{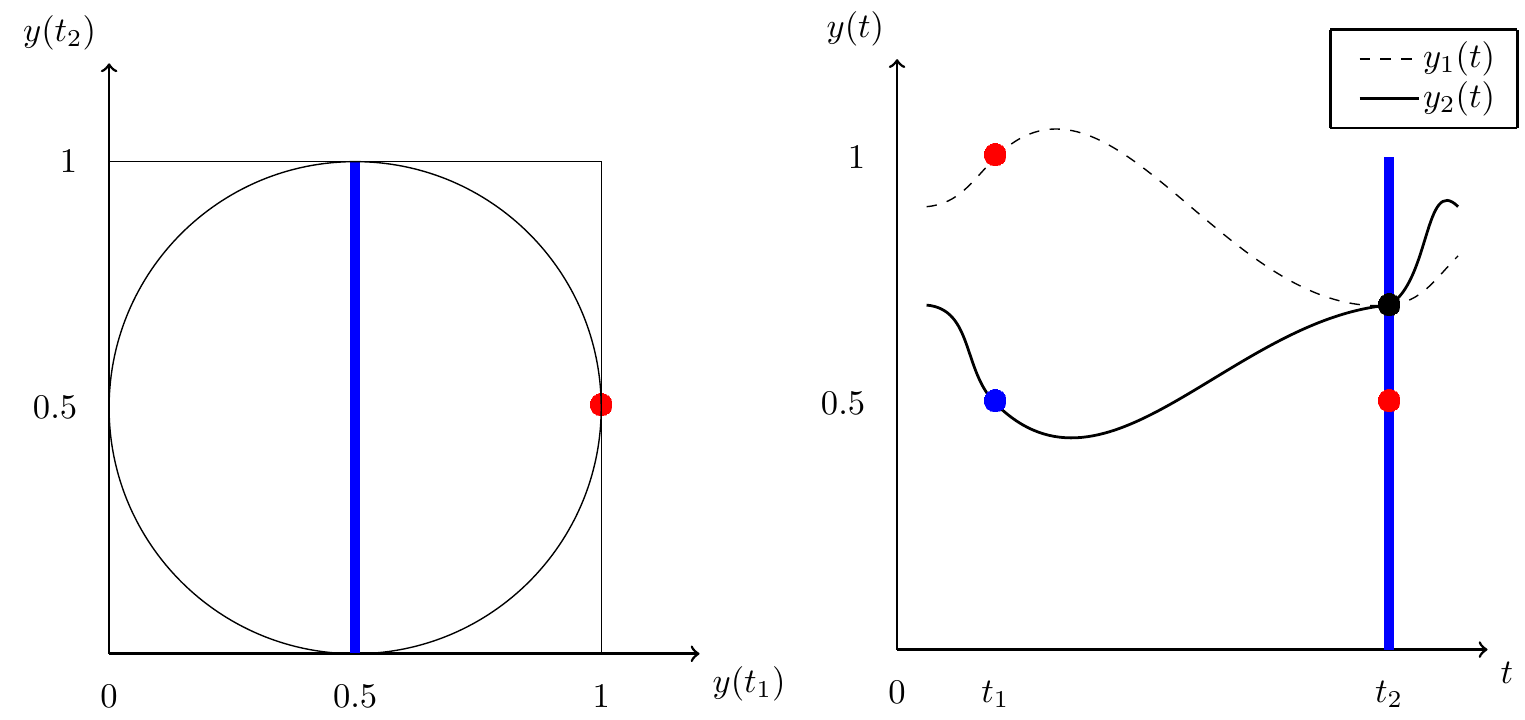}
\end{center}
\caption{Example of importance of obtaining prediction bands.}
\label{fig:IOBR}
\end{figure}
In order to clarify this concept, let us consider the following example. Let $\mathcal{C}^1_n$ be a prediction band and let us consider the simple case in which $B_n(t)$ is the interval $[0,1]$ for each $t \in \mathcal{T}$: in so doing, from a geometrical point of view $\mathcal{C}^1_n$ is an infinite-dimensional hypercube. Specifically, let us focus on two points of the domain, $t_1$ and $t_2$ respectively, and with a slight abuse of notation let us indicate with $\mathcal{C}^1_n(t_1,t_2):=\{(y(t_1),y(t_2)): (y(t_1),y(t_2)) \in [0,1] \times [0,1] \}$ the "restriction" of prediction band related to $\{t_1,t_2\}$.
In addition, let $\mathcal{C}^2_n$ be for example a different hypothetical  prediction set having the shape of an infinite-dimensional hyper-sphere such that for instance $\mathcal{C}^2_n(t_1,t_2):=\{(y(t_1),y(t_2)): (y(t_1)-0.5)^2 + (y(t_2)-0.5)^2 \leq 0.5^2 \}$, i.e. $\mathcal{C}^2_n(t_1,t_2)$ is the closed disk of center (0.5,0.5) and radius 0.5. Both $\mathcal{C}^1_n(t_1,t_2)$ and $\mathcal{C}^2_n(t_1,t_2)$ are plotted on the left side of Figure \ref{fig:IOBR}. Drawing conclusions only on the basis of the behavior of the plotted functions in $t_1$ and $t_2$  and ignoring it in all the other points of the domain, the right side of Figure \ref{fig:IOBR} shows a function that does not belong to $\mathcal{C}^2_n$ (the dashed curve $y_1$) and a function that belongs to such set (the solid curve $y_2$): indeed, conditional on the fact that $y_1(t_1)=1$, the dashed curve $y_1$ must satisfy $y_1(t_2)=0.5$ to be included in $\mathcal{C}^2_n$, as shown by the red dot on the left of Figure \ref{fig:IOBR}. Conversely, conditional on the fact that $y_2(t_1)=0.5$, $y_2(t_2)$ can assume whatever value between 0 and 1 to be included in $\mathcal{C}^2_n$, as shown by the blue solid vertical line on the left of Figure \ref{fig:IOBR}. The fact that the point where $y_1$ and $y_2$ intersect (the black dot on the right of Figure \ref{fig:IOBR}) determines whether to include or not a function in $\mathcal{C}^2_n$ on the basis of the value assumed by that function in $t_1$ represents an undeniable limit to the visualization of prediction sets, especially considering that this phenomenon involves all $t \in \mathcal{T}$. 
Fortunately, this problem is completely avoided by prediction sets as $\mathcal{C}^1_n$, and more generally by every prediction band: indeed, differently from prediction sets characterized by other shapes, prediction bands always coincide with (and are not only a subset of) their envelope. In view of this, the development of a method that necessarily outputs prediction bands - instead of more general prediction sets - represents the starting point of this work. 

\section{Conformal Prediction Bands}
\label{sec:conformal_prediction_bands}

\subsection{Conformal Prediction}
\label{sec:conformal_approach}

The framework we use to develop our prediction sets is \textit{Conformal Prediction}, a nonparametric approach proposed in the multivariate literature for the first time by \citet{gammerman_learning_1998} and thoroughly described in  \citet{vovk2005algorithmic}, that can be used to construct finite-sample either valid or exact prediction sets under no assumptions other than \textit{i.i.d.} data \citep[for a review of the topic see, e.g.,][]{lei2018distribution, zeni_conformal_2020}. Even though the theory holds also under the weaker assumption of exchangeable data, in this manuscript we will focus on the case of \textit{i.i.d.} data which is a very common case in applications and in particular on the case of \textit{i.i.d.} functional data taking value in $\mathcal{Y}(\mathcal{T})$.

Following the notation of \citet{vovk2005algorithmic}, given a set of \textit{i.i.d.} random functions  $Y_1, \ldots, Y_n \sim P$ and an independent random function $Y_{n+1}  \sim P$, a valid prediction set $\mathcal{C}_{n,1-\alpha}:=\mathcal{C}_{n,1-\alpha}(Y_1,\dots,Y_n)$ for  $Y_{n+1}$ is a set such that

\begin{equation}
\mathbb{P} \left( Y_{n+1} \in \mathcal{C}_{n,1-\alpha} \right) \geq 1-\alpha
\label{eq:cons_valid}
\end{equation}
for any significance level $\alpha \in (0,1)$  and with $\mathbb{P}$ the probability corresponding to the product measure induced by  $P$ \citep{lei2015conformal}. 
If the inequality in (\ref{eq:cons_valid}) is replaced by the equality, the prediction set is also said to be exact. In order to avoid ambiguity, later in the discussion the term \textit{coverage} (or \textit{unconditional coverage}) will be used to refer to $\mathbb{P} \left( Y_{n+1} \in \mathcal{C}_{n,1-\alpha} \right)$, the term \textit{conditional coverage} will be used to refer to $\mathbb{P} \left( Y_{n+1} \in \mathcal{C}_{n,1-\alpha} | \mathcal{C}_{n,1-\alpha}  \right)$ and the terms \textit{empirical coverage} and \textit{empirical conditional coverage} will be used to refer to the estimate - from simulated data - of the coverage and conditional coverage respectively.

 Specifically, we will focus on the Semi-Off-Line Inductive Conformal framework, also known simply as \textit{Split Conformal}, a computationally efficient modification of the original Transductive Conformal method  \citep[firstly proposed in][]{papadopoulos2002inductive}. In order to present this approach, let us consider the following procedure: given data $y_1, \ldots, y_n$, let $\{1,\dots, n\}$ be randomly divided into two sets $\mathcal{I}_1, \mathcal{I}_2$ and let us define the training set as $\{ y_h : h \in \mathcal{I}_1\}$ and the calibration set as $\{ y_h : h \in \mathcal{I}_2\}$, with $\vert \mathcal{I}_1 \vert=m$, $\vert \mathcal{I}_2 \vert=l$ and  $m,l \in \mathbb{N}_{>0}$ such that $n=m+l$. Let us also define \textit{nonconformity measure} as any measurable function $A(\{y_h: h \in  \mathcal{I}_1 \} ,y)$ taking values in $\bar{\mathbb{R}}$ whose aim is to score how different $y \in \mathcal{Y(T)}$ is from the training set. 
 The split conformal prediction set constructed on the basis of the observed sample $y_1,\dots,y_n$ is defined as $\mathcal{C}_{n, 1-\alpha}:= \left\{y \in \mathcal{Y(T)}: \delta_{y}>\alpha \right\}$, with 
\begin{equation*}
\delta_y :=  \frac{\left|\left\{j \in  \mathcal{I}_2 \cup \{n+1\} : R_{j} \geq R_{n+1}\right\}\right|}{l+1}
\end{equation*}
and \textit{nonconformity scores} $R_j:=A( \{y_h: h \in  \mathcal{I}_1 \} ,y_{j})$ for $j \in \mathcal{I}_2$, $R_{n+1}:=A( \{y_h: h \in  \mathcal{I}_1 \} ,y)$. In particular, hereafter we will focus on nonconformity scores $\{ R_h : h \in \mathcal{I}_2\}$ having a continuous joint distribution, an assumption generally satisfied in the functional context.

The essential result \citep[due to][]{vovk2005algorithmic} traditionally evoked when dealing with the Conformal approach concerns the validity of split prediction sets: indeed, under the exchangeability assumption (a direct consequence of having i.i.d. data) $\delta_Y$ is uniformly distributed over $\{ 1/(l+1), 2/(l+1), \dots, 1\}$ and then (\ref{eq:cons_valid}) holds.  Theorem \ref{th_conservativeness_up_low_bound} proves and enriches such known result by making the exact probability reached by split prediction sets explicit. The proof is given in Appendix A.1.

\begin{theorem}
Let $\mathcal{C}_{n, 1-\alpha}$ be a split conformal prediction set. If $Y_1,\dots,Y_{n+1}$ are i.i.d. and $\{ R_h : h \in \mathcal{I}_2\}$ have a continuous joint distribution, then 
\begin{equation*}
\mathbb{P} \left( Y_{n+1} \in \mathcal{C}_{n,1-\alpha} \right) =1-\frac{\lfloor (l+1)\alpha \rfloor}{l+1}.
\end{equation*}
Specifically, $\mathcal{C}_{n, 1-\alpha}$ always satisfies 
\begin{equation}
1-\alpha \leq \mathbb{P} \left( Y_{n+1} \in \mathcal{C}_{n,1-\alpha} \right) < 1 - \alpha + \frac{1}{l+1}.
\label{eq_upper_bound_lei}
\end{equation}

\label{th_conservativeness_up_low_bound}
\end{theorem}

A natural consequence of the first part of Theorem \ref{th_conservativeness_up_low_bound} is that when $\lfloor (l+1)\alpha \rfloor = (l+1)\alpha$ the procedure automatically outputs exact prediction sets: in practice, since in most cases both $\alpha$ and $l$ are given by the application in hand, such property should be simply considered as an useful by-product that may occur in some circumstances. More generally, Theorem \ref{th_conservativeness_up_low_bound} states that Conformal approach ensures an easy-to-compute precise coverage for split prediction sets, and not only their validity.  Furthermore, the second part of Theorem \ref{th_conservativeness_up_low_bound} suggests that the coverage provided by split conformal prediction sets is no less than $1-\alpha$ and over-coverage is basically avoided when sample size is large. In particular, inequality (\ref{eq_upper_bound_lei}) represents a minimal modification of Theorem 2 of \citet{lei2018distribution}: the only difference - besides notation - is the change of `'$\leq$'' with ``$<$'' in the upper bound of (\ref{eq_upper_bound_lei}).

Conformal inference is a field of deep interest as minimal assumptions are required on $P$ to obtain prediction sets satisfying (\ref{eq:cons_valid}) for any finite sample size $n$, a property particularly appealing in the functional context. 
A slight modification \citep{vovk2005algorithmic} of the aforementioned procedure even allows to obtain a stronger version of Theorem \ref{th_conservativeness_up_low_bound}: in order to present it, first of all let us introduce an element of randomization $\tau_{n+1}$, realization of a uniform random variable in $[0,1]$. 
The smoothed split conformal prediction set is defined as $\mathcal{C}_{n,1-\alpha,\tau_{n+1}}:=\left\{y \in \mathcal{Y(T)}: \delta_{y,\tau_{n+1}}>\alpha \right\}$, with 
\begin{equation*}
\delta_{y,\tau_{n+1}} := \frac{\left|\left\{j \in  \mathcal{I}_2: R_{j} > R_{n+1}\right\}\right| + \tau_{n+1} \left|\left\{j \in  \mathcal{I}_2 \cup \{n+1\}: R_{j} = R_{n+1}\right\}\right|}{l+1}.
\end{equation*}
Smoothed split conformal prediction sets are, by construction, exact for any $\alpha,l$, i.e. $\mathbb{P} \left( Y_{n+1} \in \mathcal{C}_{n,1-\alpha,\tau_{n+1}} \right)= 1 - \alpha$: to the best of our knowledge, in the literature there is no formal proof of this well-established result \citep[due to][]{vovk2005algorithmic}, and so a 
proof is given in Appendix A.1.

\begin{remark}
Our discussion was limited to the split setting because our work only focuses on it, but the results of this section are very general and require just little changes to be applied to the Transductive/Full Conformal framework. In addition, as highlighted by \citet{vovk2005algorithmic} and briefly mentioned at the beginning of this section, Theorem \ref{th_conservativeness_up_low_bound} and the result about exactness of smoothed prediction sets hold even when the weaker assumption of exchangeability is formulated instead of the traditional hypothesis of i.i.d. data.
\end{remark}

\begin{remark}
The division of data into the training and calibration sets always induces an element of randomness  into the procedure, also in the non-smoothed scenario. A possible approach to limit the effect of this evidence consists of combining prediction sets obtained from different splits, but the results provided by \citet{lei2018distribution} suggest to perform a single split. As a consequence, in this article the aforementioned single-split process is considered.
\end{remark}

\begin{remark}
The Conformal approach can be also successfully applied to regression and classification problems. A detailed presentation is not included hereafter being out of scope, but an exhaustive discussion can be found in \citet{vovk2005algorithmic}.
\end{remark}

\begin{remark}
Although we focus on the functional setting, the Conformal framework has initially been developed in the traditional univariate and multivariate settings and so all arguments and results presented in this section can also be applied to univariate variables and random vectors.
\end{remark}

\subsection{Supremum Metric as a Nonconformity Measure}
\label{sec:supremum_metric}

Although some authors proposed different approaches to find prediction bands under the Gaussian assumption \citep{yao2005functional} and through finite dimensional projection \citep{lei2015conformal}, to the best of our knowledge no method to create valid prediction bands by only assuming i.i.d. functional data and by avoiding dimension reduction is available in the literature. 

 In light of this and of the discussion in Section \ref{sec:importance_rectangular}, we propose a fast and scalable split conformal predictor that outputs closed-form  finite-sample valid (or even exact) prediction bands under only the i.i.d. assumption.
Indeed, the Conformal framework ensures, by construction, that the prediction sets obtained are always valid, but other features such as shape and size depend on the specific nonconformity measure used: as a consequence, the core of the Conformal approach is represented by the choice of such measure.  

In particular, the nonconformity measure we propose automatically allows to obtain prediction bands and is based on the supremum metric:
\begin{equation}
A(\{y_h: h \in  \mathcal{I}_1 \}, y)=\sup_{t \in \mathcal{T}} \left| y(t)-g_{\mathcal{I}_1}(t)\right|,
\label{eq:original_NCM}
\end{equation}
with  $g_{\mathcal{I}_1}: \mathcal{T} \rightarrow \mathbb{R}$ a function belonging to $L^{\infty}(\mathcal{T})$ based on  $\{y_h: h \in  \mathcal{I}_1 \}$ and acting as a point predictor of the new observation. Given the assumptions on $\mathcal{Y(T)}$ and $\mathcal{T}$, the computation of the supremum in (\ref{eq:original_NCM}) could be replaced by the simpler computation of the maximum: however, for historical reasons and consistency with possible future developments of the current work, we will use the standard notation (\ref{eq:original_NCM}). Although valid prediction bands are obtained regardless the specific $g_{\mathcal{I}_1}$ involved, a careful choice of this function helps to obtain small prediction bands, a desirable property from an application point of view which will be investigated in Section \ref{sec:modulation}  \citep{lei2018distribution}. In view of this, $g_{\mathcal{I}_1}$ is typically a point predictor summarizing information provided by $\{y_h: h \in  \mathcal{I}_1 \}$, e.g. the sample functional mean. 
However, since the purpose of the article is to construct either valid or exact prediction bands starting from any point predictor in order to obtain a widely usable procedure, later in the discussion we will always consider $g_{\mathcal{I}_1}$ as given - and properly chosen by the expert according to the specific framework considered.
Focusing on the non-smoothed scenario (the minor changes needed for the smoothed case are introduced in Appendix A.4), first of all it is possible to notice that if $\alpha \in (0,1/(l+1) )$ then $\mathcal{C}_{n, 1-\alpha}= \mathcal{Y(T)}$ since $\delta_y$ can not be less than $1/(l+1)$: for this reason, later in the discussion we will always consider $\alpha \in [1/(l+1),1 )$, unless otherwise stated. If $\alpha \in [1/(l+1),1 )$, the definition of $\mathcal{C}_{n, 1-\alpha}$ and $\delta_y$ implies that $y \in \mathcal{C}_{n, 1-\alpha} \iff R_{n+1} \leq k$, with $k$ the $\lceil (l+1)(1-\alpha) \rceil$th smallest value in the set $\{ R_h: h \in \mathcal{I}_2 \}$. Then 
\begin{align*}
\sup_{t \in \mathcal{T}} \left| y(t)-g_{\mathcal{I}_1}(t)\right| \leq k 
&\iff  \left| y(t)-g_{\mathcal{I}_1}(t)\right| \leq k \quad \forall t \in \mathcal{T}  \\
&\iff y(t) \in [g_{\mathcal{I}_1}(t)-k, g_{\mathcal{I}_1}(t)+k] \quad \forall t \in \mathcal{T}.
\end{align*}
Therefore, the split conformal prediction set induced by the nonconformity measure (\ref{eq:original_NCM}) is
\begin{equation}
\mathcal{C}_{n, 1-\alpha}:= \left\{y \in \mathcal{Y(T)}: y(t) \in [g_{\mathcal{I}_1}(t)-k, g_{\mathcal{I}_1}(t)+k] \quad \forall t \in \mathcal{T} \right\}.
\label{eq:pred_set_original}
\end{equation}

Besides having the shape of a a band, the introduced prediction set can be found in closed form, an appealing property that incredibly speeds up computation time. In addition, the Conformal framework and the simplicity of the nonconformity measure ensure highly scalable prediction bands as, on top of the cost needed to build the point predictor $g_{\mathcal{I}_1}$, the time required to find $k$ increases linearly with $l$. Then, if a particularly sophisticated predictor is chosen for $g_{\mathcal{I}_1}$, one is justified in expecting the total computation cost to be dominated by the calculation of such point predictor. 
Moreover, as usual in the prediction framework the band is built around a ``central'' object ($g_{\mathcal{I}_1}$ in this case), a fact that further suggests to define this function as a data-driven point predictor. Finally, the prediction bands defined in (\ref{eq:pred_set_original}) are simultaneous by construction, i.e. bands ensuring the desired coverage globally (in addition to the pointwise validity). Similarly to the multivariate setting, a simple concatenation of pointwise prediction intervals based on the pointwise nonconformity score $\left| y(t)-g_{\mathcal{I}_1}(t)\right|$ for all $t \in \mathcal{T}$ would lead to a prediction band: that is a subset of the simultaneous prediction band (\ref{eq:pred_set_original}) (the proof is given in Appendix A.2); with guaranteed pointwise coverage for all $t \in \mathcal{T}$; but whose simultaneous coverage over the domain $\mathcal{T}$ can be dramatically lower than the desired one.

\subsection{Improving Efficiency: the Choice of the Modulation Function}
\label{sec:modulation}

\begin{figure}
\begin{center}
\includegraphics[width=14.75cm,height=3.958cm]{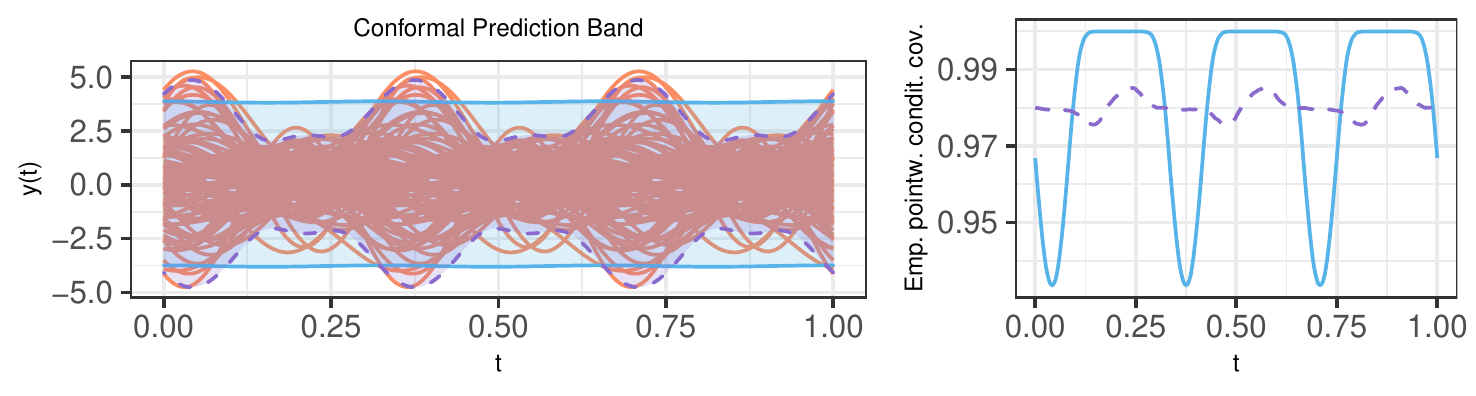}
\end{center}
\caption{The left panel shows the split conformal prediction band computed as in (\ref{eq:pred_set_original}) (solid light blue band) and that computed as in (\ref{eq:pred_set_modulated}) by considering the standard deviation function as $s_{\mathcal{I}_1}$ (dashed purple band). For visualization, a random subsample of $y_1,\dots,y_{198}$ is plotted. The right panel shows the empirical pointwise conditional coverage reached by the first band (solid light blue line) and by the second one (dashed purple line). $\alpha=0.1$. \label{func_band_original_vs_standardized_plus_empcov}}
\end{figure}

It can be easily noted that the width of (\ref{eq:pred_set_original}) over $\mathcal{T}$ is constant and equal to $2k$ but, intuitively, prediction bands that do not adapt their width according to the local variability of functional data, even though theoretically sound,  may be of limited interest in real applications.  
Let us consider the following running example: let $y_1,\dots,y_{198}$ be independent realizations of the random function $Y(t):= X_1 + X_2 \cos(6\pi t) + X_3 \sin(6\pi t)$, with $t \in [0,1]$ and $(X_1, X_2, X_3)$ being a Gaussian random vector such that $\mathrm{E}[X_i]=0$, $\mathrm{Var}[X_i]=1$,  $\mathrm{Cov}[X_i, X_j]=0.6$ for $i,j=1,2,3$, $i \neq j$. The solid light blue band in the left panel of Figure \ref{func_band_original_vs_standardized_plus_empcov}
shows the prediction band obtained by the procedure presented in Section \ref{sec:supremum_metric} considering $\alpha=0.1$, $m=n/2$ and $g_{\mathcal{I}_1}$ sample functional mean of the training set: given the different variability of functional data over $\mathcal{T}$, in the low-variance parts of the domain the prediction band is dramatically large containing all the pointwise evaluations of the functional data (see, for example, $t=0.5$ and nearby points).

A possible solution to this drawback consists of defining the following nonconformity measure and nonconformity scores:
\begin{equation}
A( \{y_h: h \in  \mathcal{I}_1 \}, y)= \sup_{t \in \mathcal{T}} \left| \frac{y(t)-g_{\mathcal{I}_1}(t)}{s_{\mathcal{I}_1}(t)}\right|,
\label{eq:modulated_NCM}
\end{equation}
\begin{equation*}
R^{s}_j:=\sup_{t \in \mathcal{T}} \left| \frac{y_{j}(t)-g_{\mathcal{I}_1}(t)}{s_{\mathcal{I}_1}(t)}\right|, \quad  R^{s}_{n+1}:=\sup_{t \in \mathcal{T}} \left| \frac{y(t)-g_{\mathcal{I}_1}(t)}{s_{\mathcal{I}_1}(t)}\right|,
\end{equation*}
with $j \in \mathcal{I}_2$ and $s_{\mathcal{I}_1}:=s(\{y_h: h \in  \mathcal{I}_1 \}): \mathcal{T} \rightarrow \mathbb{R}_{>0}$ a function which belongs to $L^{\infty}(\mathcal{T})$ based on  $\{y_h: h \in  \mathcal{I}_1 \}$. At the interpretative level, the new nonconformity measure (\ref{eq:modulated_NCM}) can be suitably considered as the nonconformity measure (\ref{eq:original_NCM}) taking the transformed functions $y^{s}(t):= y(t) / s_{\mathcal{I}_1}(t)$ and $g^{s}_{\mathcal{I}_1}(t)=g_{\mathcal{I}_1}(t)/s_{\mathcal{I}_1}(t)$ $\forall t \in \mathcal{T}$ as input instead of the original functions $y(t)$, $g_{\mathcal{I}_1}(t)$. It is important to notice that, since $s_{\mathcal{I}_1}(t)>0$ $\forall t \in \mathcal{T}$, the function $s_{\mathcal{I}_1}$ modulates the original data without altering the order of the functions at each point $t$: for this reason, later in the discussion the term \textit{modulation function} will be used to refer to $s_{\mathcal{I}_1}$.

Therefore, the  split conformal prediction band induced by the nonconformity measure (\ref{eq:modulated_NCM}), obtained by replicating the computations of Section \ref{sec:supremum_metric} (see Appendix A.3 for the proof), is
\begin{equation}
\mathcal{C}^s_{n, 1-\alpha}:= \left\{y \in \mathcal{Y(T)}: y(t) \in [g_{\mathcal{I}_1}(t)-k^s s_{\mathcal{I}_1}(t) , g_{\mathcal{I}_1}(t)+k^s s_{\mathcal{I}_1}(t)] \quad \forall t \in \mathcal{T}  \right\},
\label{eq:pred_set_modulated}
\end{equation}
with  $k^s$ the $\lceil (l+1)(1-\alpha) \rceil$th smallest value in the set $\{ R^{s}_h: h \in \mathcal{I}_2 \}$. In other words, the procedure presented in this section consists of modulating the data, computing the prediction band (\ref{eq:pred_set_original}) by using the transformed data and back-transforming it in the non-modulated space: in so doing, prediction bands adapt their width according to the specific modulation function chosen and their validity 
is guaranteed by the Conformal framework. A similar consideration has been highlighted also in the scalar regression setting by \citet{lei2018distribution}, who proposed a locally weighted Split Conformal method to vary the width of the prediction sets over the covariates $x \in \mathbb{R}^p$.

In order to understand the modification introduced by the modulation function, let us consider the aforementioned running example and specifically the left panel of Figure \ref{func_band_original_vs_standardized_plus_empcov}: in this case, the band obtained by considering the standard deviation function
\citep{ramsay_functional_2005} as $s_{\mathcal{I}_1}$ (dashed purple band) is deeply different from the one in the top panel and it seems to better adapt to the variability of the data over $\mathcal{T}$. Intuitively, one is justified in accepting the bands to become wider in the parts of the domain where data show high variability  in order to obtain narrower and more informative prediction bands in those parts characterized by low variability. 

\begin{remark}
Replacing function $s_{\mathcal{I}_1}$ with $s_{\mathcal{I}_2}$ does not allow to obtain closed-form valid prediction bands. This is due to the fact that their dependence on the calibration set involves $\{R^{s}_h: h \in \mathcal{I}_2  \cup \{n+1\} \}$ not being exchangeable, and consequently validity not being guaranteed.
\end{remark}


\begin{remark}
\label{remark_integral_equal_1}
Prediction bands induced by the modulation functions $s_{\mathcal{I}_1}$ and $\lambda \cdot s_{\mathcal{I}_1}$, with $\lambda \in \mathbb{R}_{>0}$, are identical. The proof is given in Appendix A.3. As a consequence, an equivalence relation naturally arises and so for each specific equivalence class (made up of modulation functions equal up to a multiplicative factor) we will consider the modulation function whose integral is equal to 1. In view of this, the original nonconformity measure (\ref{eq:original_NCM}) can be interpreted as the nonconformity measure induced by the modulation function $s^0(t):=1/ \vert \mathcal{T} \vert$ $\forall t \in \mathcal{T}$, whose notation does not include the subscript $\mathcal{I}_1$ to underline the lack of dependence of this function on the training set. 
\end{remark}

\begin{remark}
One of the aim of the introduction of $ s_{\mathcal{I}_1}$ is to reduce the variability of the pointwise miscoverage over $\mathcal{T}$. In order to clarify this concept, let us consider the right panel of Figure \ref{func_band_original_vs_standardized_plus_empcov}. The solid light blue (dashed purple respectively) line shows the empirical pointwise conditional coverage of the solid light blue (dashed purple respectively) prediction band showed in the left panel of the same figure, that was obtained by setting $\alpha=0.1$. The empirical conditional  coverage has been computed considering the number of times that 200,000 - independent from and identically distributed to the original sample - new functions belong to the two prediction bands over $\mathcal{T}$.
As expected, the absence of modularization involves the empirical pointwise converage being highly variable over $\mathcal{T}$, whereas the use of the standard deviation function as modulation function leads to an empirical pointwise coverage concentrated around 0.98. 
\end{remark}

However, in absence of an optimality criterion there are no formal reasons to prefer a specific modulation function over another, as Conformal approach ensures valid prediction sets regardless the choice of $s_{\mathcal{I}_1}$. In this regard, a criterion that naturally arises in the prediction framework to discriminate between modulation functions is maximization of efficiency, i.e. minimization of the size of prediction sets \citep{vovk2005algorithmic} . The reason of this choice is very intuitive: since prediction bands are, by construction, valid, one is justified in seeking small prediction bands because they include subregions of the sample space where the probability mass is concentrated \citep{lei2013distribution}. In view of this, first of all it is essential to define what the size of a prediction band is, a nontrivial topic in the functional framework. The definition we will consider is simply the area between the upper and lower bound of the prediction band:
\begin{equation}
\mathcal{Q}(s_{\mathcal{I}_1}):=\int_{\mathcal{T}} 2 \cdot k^s \cdot s_{\mathcal{I}_1}(t) dt =2 \cdot k^s,
\label{eq:area}
\end{equation}
that is equal to $k^s$ up to a constant and proportional to $2 k^s/|\mathcal{T}|$, i.e. the average width of the prediction band over the domain $\mathcal{T}$. 

Formally, in the usual finite-dimensional setting the aim would be to find the optimal modulation function that minimizes the risk functional $\mathrm{E}[k^s]$. Unfortunately, in the functional setting even the concept of probability density function is generally not well defined since there is no $\sigma$-finite dominating measure \citep{delaigle2010defining}, 
and so that minimization is not feasible for general $P$. As a consequence, the minimization problem must be simplified: by considering $k^s$ as a non-random quantity depending on observed functions $y_1,\dots,y_n$ instead of random functions $Y_1,\dots,Y_n$, the aim becomes the direct minimization of $k^s$. Although initially it may seem like an oversimplification to some readers, it is important to underline that this approach is made possible by a well-established principle representing the core idea of many algorithms and methods  (e.g. machine learning techniques)  known as empirical risk minimization principle \citep{vapnik1992principles}. 

The proposed adjustment reduces the complexity of the optimization task, but the problem still presents tricky aspects. Indeed, not only the minimization can not be analytically addressed by calculus of variations given the complexity of $k^s$, 
but also the optimal modulation function can not be uniquely determined given the specific structure of  $ R_h^s, h \in \mathcal{I}_2$. In fact, the dependency of $s_{\mathcal{I}_1}$ only on the functions of the training set and of the numerator of $R_h^s$ (i.e. $ \left| y_{h}(t)-g_{\mathcal{I}_1}(t) \right|, \quad h \in \mathcal{I}_2$) also on the functions of the calibration set makes the optimization unfeasible for all $P$ and the general problem ill-posed.

In such a non-standard context, the line of reasoning must necessarily be changed. Therefore, in the discussion below we focus on finding a function - called c-function hereafter for the sake of simplicity -  satisfying the definition of modulation function but depending also on the calibration set through $ \{ y_h : h \in \mathcal{I}_2\} $ and such that
\begin{enumerate}
\item For $m,l \to +\infty$ it converges to a given function and its training counterpart (i.e. the function - called t-function hereafter - equal to the c-function but whose dependence on $ \{ y_h : h \in \mathcal{I}_2\}$ is replaced by the dependence on the training set through $ \{ y_h : h \in \mathcal{I}_1\})$ converges to the same function
\item it leads to prediction bands that are not wider (in the sense of (\ref{eq:area})) than those obtained by not modulating (i.e. by using $s^0$)
\end{enumerate}
If these two conditions are met, the use of the t-function as modulation function ensures that valid prediction bands are obtained (due to its dependence only on $\{ y_h : h \in \mathcal{I}_1\})$ and that asymptotically the second condition is satisfied.
Specifically, that condition represents a desirable and appealing property since, if violated, the modulation process could represent a meaningless complication compared to the original nonconformity measure (\ref{eq:original_NCM}).

In order to construct a c-function able to meet these two conditions, it is important to focus on what $k^s$ is: ignoring just for now the contribution of the modulation function, $k^s$ is a quantity derived by the $\lceil (l+1)(1-\alpha) \rceil$th least extreme function between those in the calibration set, in which the concept of "extreme" is naturally induced by the supremum metric. In light of this, the guidelines we decided to follow in the construction of a meaningful c-function are two. First of all, the behavior of the $l-\lceil (l+1)(1-\alpha) \rceil$ most extreme functions in the calibration set should not be taken into account since they do not affect the value of $k^s$. Secondly, given the specific nonconformity measure considered, the c-function should modulate data considering the remaining $\lceil (l+1)(1-\alpha) \rceil$ functions on the basis of the most extreme value observed $\forall t \in \mathcal{T}$.

Inspired by these guidelines, we propose the following c-function:
\begin{equation}
 \bar{s}^{c}_{\mathcal{I}_1}(t):=\frac{\max_{j \in \mathcal{H}_2} |y_{j}(t)- g_{\mathcal{I}_1}(t)|}{\int_{\mathcal{T}} \max_{j \in \mathcal{H}_2} |y_{j}(t)- g_{\mathcal{I}_1}(t)| dt}
\label{s_calibrationdef}
\end{equation}
with 
\begin{equation*}
\mathcal{H}_2:=\{j \in \mathcal{I}_2: \sup_{t \in \mathcal{T}} |y_{j}(t)-g_{\mathcal{I}_1}(t)| \leq k \} 
\end{equation*}
and $k$ defined as in Section \ref{sec:supremum_metric}, i.e. the $\lceil (l+1)(1-\alpha) \rceil$th smallest value in the set $\{ R_h: h \in \mathcal{I}_2 \}$. 
The corresponding t-function is
\begin{equation}
 \bar{s}_{\mathcal{I}_1}(t):=\frac{\max_{j \in \mathcal{H}_1} |y_{j}(t)- g_{\mathcal{I}_1}(t)|}{\int_{\mathcal{T}} \max_{j \in \mathcal{H}_1} |y_{j}(t)- g_{\mathcal{I}_1}(t)| dt}
\label{s_training}
\end{equation}
with $\mathcal{H}_1=\mathcal{I}_1$ if $\lceil (m+1)(1-\alpha) \rceil > m$, otherwise
\begin{equation*}
\mathcal{H}_1:=\{j \in \mathcal{I}_1: \sup_{t \in \mathcal{T}} |y_{j}(t)-g_{\mathcal{I}_1}(t)| \leq \gamma \} 
\end{equation*}
with $\gamma$ the $\lceil (m+1)(1-\alpha) \rceil$th smallest value in the set $\{ \sup_{t \in \mathcal{T}} |y_{h}(t)-g_{\mathcal{I}_1}(t)|: h \in \mathcal{I}_1 \}$.

In order not to overcomplicate the notation, in the definition of $ \bar{s}^{c}_{\mathcal{I}_1}$ and $ \bar{s}_{\mathcal{I}_1}$ we quietly assumed that both numerators are different from 0 $\forall t \in \mathcal{T}$ almost surely. If not, the adjustment described in Appendix A.3 is developed.
From an operational point of view, t-function $\bar{s}_{\mathcal{I}_1}(t)$ ignores the most extreme functions (i.e. the functions belonging to $\mathcal{I}_1 \setminus \mathcal{H}_1$)  and modulates
data on the basis of the remaining non-extreme functions. Specifically, the dependence of $\gamma$ on $\alpha$ allows to provide carefully chosen modulation process according to the specific level $1-\alpha$ chosen for the prediction set.

The fulfillment of the two aforementioned conditions by the function (\ref{s_calibrationdef}) is proved by the following two theorems. 

\begin{theorem}
\label{th:shared_convergence}
 Let $m/n = \theta$ with $0 < \theta < 1$ and let 
$\mathrm{Var}[g_{\mathcal{I}_1}(t)] \to 0 $ when $m \to +\infty$. Then  $\bar{s}^{c}_{\mathcal{I}_1}$ and $ \bar{s}_{\mathcal{I}_1}$ converge to the same function when $n \to +\infty$ and $\lim_{n \to +\infty} \mathcal{C}^{\bar{s}}_{n, 1-\alpha}= \lim_{n \to +\infty} \mathcal{C}^{\bar{s}^c}_{n, 1-\alpha}$ $\forall$ $\alpha \in (0,1)$.
\end{theorem}

\begin{theorem}
\label{th:better_than_not_modul}
$\mathcal{Q}(s^{0}) \geq \mathcal{Q}( \bar{s}^{c}_{\mathcal{I}_1})$. Specifically, $\mathcal{Q}(s^{0}) = \mathcal{Q}( \bar{s}^{c}_{\mathcal{I}_1})$ if and only if $\max_{j \in \mathcal{H}_2} |y_{j}(t)- g_{\mathcal{I}_1}(t)|$ \text{is constant almost everywhere}.
\end{theorem}

Both proofs are given in Appendix A.3. 
It is important to notice that Theorem \ref{th:shared_convergence} requires very mild conditions, an evidence that allows it to hold in many general contexts. 

In light of this,
 the function (\ref{s_training}) represents an outstanding candidate in the choice of the modulation function since the Conformal setting and the nonconformity measure (\ref{eq:modulated_NCM}) guarantee valid prediction bands - as well as all the other desirable properties highlighted in Section \ref{sec:supremum_metric} - and at the same time to asymptotically obtain prediction bands no less efficient than those induced by $s^{0}$.

\begin{remark}

The fact that $\bar{s}^{c}_{\mathcal{I}_1}(t)$  leads to prediction bands that are not wider than those obtained by not modulating is not the only relevant result that is possible to obtain.  The following Theorem shows 
that prediction bands induced by $\bar{s}^c_{\mathcal{I}_1}$ are also smaller than those induced by the functions belonging to a specific group. This theorem provides a further theoretical justification for preferring  function (\ref{s_training}) to other possible modulation functions.

\begin{theorem}
\label{th:better_than_group_of_functions}
Let us define  $\mathcal{CH}_2:=\mathcal{I}_2 \setminus \mathcal{H}_2$ and let $t^{*}_i$ be the value such that 
\begin{equation} 
|y_{i}(t^{*}_i)- g_{\mathcal{I}_1}(t^{*}_i)|= \sup_{t \in \mathcal{T}} |y_{i}(t)- g_{\mathcal{I}_1}(t)| \quad \forall i \in \mathcal{I}_2.
\label{dim2tstar}
\end{equation} 
If $t^*_i$ is not unique, it is randomly chosen from the values that satisfy \eqref{dim2tstar}.

Let $s^d_{\mathcal{I}_1}$ be a modulation function such that:
\begin{enumerate}
\item $s^d_{\mathcal{I}_1} \neq \bar{s}^c_{\mathcal{I}_1} $ in the sense of Lebesgue, \textit{i.e.} $\exists$  $\mathcal{T}^* \subseteq \mathcal{T}$ such that  $s^d_{\mathcal{I}_1}(t) \neq \bar{s}^c_{\mathcal{I}_1}(t) $ $\forall t \in \mathcal{T}^*$ and $\mu(\mathcal{T}^*)>0$, with $\mu$ the Lebesgue measure
\item $s^d_{\mathcal{I}_1}(t^{*}_i) \leq \bar{s}^c_{\mathcal{I}_1}(t^{*}_i) $ $\forall i \in \mathcal{CH}_2$
\end{enumerate}   
If $\vert \mathcal{H}_2 \vert = \lceil (l+1)(1-\alpha) \rceil $, then $\mathcal{Q}(s^{d}_{\mathcal{I}_1}) > \mathcal{Q}( \bar{s}^{c}_{\mathcal{I}_1})$.
\end{theorem}

The proof is given in Appendix A.3, along with the demonstration that Theorem \ref{th:better_than_not_modul} is not a direct consequence of Theorem \ref{th:better_than_group_of_functions} since $s^{0}$ may not fulfill $s^0(t^{*}_i) \leq \bar{s}^c_{\mathcal{I}_1}(t^{*}_i) $ $\forall i \in \mathcal{CH}_2$. 
Also in this case, the field of application of Theorem \ref{th:better_than_group_of_functions} is particularly wide since the condition about the cardinality of $\vert \mathcal{H}_2 \vert$  is always met  under the assumption concerning the continuous joint distribution of $\{ R_h : h \in \mathcal{I}_2\}$ made in Section \ref{sec:conformal_approach}. 
\end{remark}

\begin{remark}
\label{remark:smoothed1}
The definitions of functions $\eqref{s_calibrationdef}$, $\eqref{s_training}$ and Theorems \ref{th:shared_convergence}, \ref{th:better_than_not_modul} and \ref{th:better_than_group_of_functions} can be easily generalized to hold also in the Smoothed Conformal framework. Technical details are provided in Appendix A.4.
\end{remark}

\section{Simulation Study}
\label{sec:simulation_study}

\subsection{Study Design}
\label{sec:sim_study_design}

In this section, we summarize the results of a two-stage simulation study comparing our approach with four alternative methods from the literature that will be detailed in the following: Naive, Band Depth, Modified Band Depth, and Bootstrap. In Section \ref{sec:sim_emp_cov} the empirical coverage is evaluated for each approach in three different scenarios, whereas in Section \ref{sec:sim_eff_study} the prediction bands obtained by the methods that guarantee a proper coverage are compared in terms of efficiency.
The hierarchical structure of the simulation study reflects the ``nested'' nature of the two features we are considering, i.e. coverage and size: indeed, the size of a prediction set should be investigated only after verifying that the method which outputted that specific prediction set guarantees the desired coverage, which represents the primary aspect when assessing prediction sets.

\begin{figure}
\begin{center}
\includegraphics[width=14.75cm,height=5.992cm]{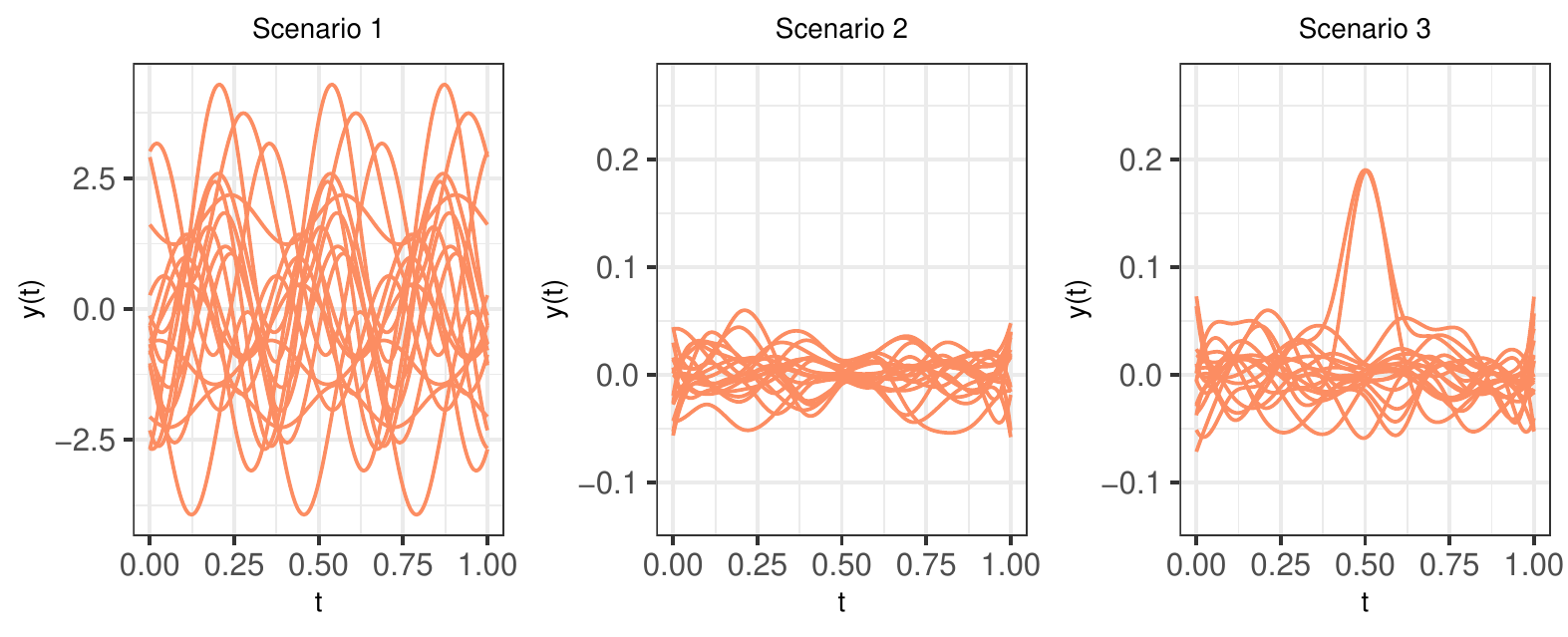}
\end{center}
\caption{Graphical representation of the scenarios. The sample size is $n=18$. }
\label{fig:scenarios}
\end{figure}

Specifically, the three scenarios allow to compare the methods in three different frameworks: when data show a constant variability over the domain (Scenario 1), when data show a different variability over the domain (Scenario 2) and when data are characterized by outliers (Scenario 3). Formally, the three scenarios are: 
\begin{itemize}
\item Scenario 1. $\forall i=1,\dots,n$
\begin{equation*}
y_i(t)= x_{i1} + x_{i2} \cos(6\pi (\,t + u_i) ) + x_{i3} \sin(6\pi \,(t + u_i))
\end{equation*}
with $\mathcal{T}=[0,1]$, $(x_{11},x_{12},x_{13})^T,\dots,(x_{n1},x_{n2},x_{n3})^T$ i.i.d. realizations of
\begin{equation*}
X \sim N_3\left(\textbf{0},\left[\begin{smallmatrix}
  1 & 0.6 & 0.6\\ 
  0.6 & 1 & 0.6\\ 
    0.6 & 0.6 & 1\\
\end{smallmatrix}\right]\right)
\end{equation*}
and $u_1,\dots, u_n$ i.i.d. realizations of 
\begin{equation*}
U \sim \text{Unif}\left[-\frac{1}{6}, \frac{1}{6}\right].
\end{equation*}

\item Scenario 2. $\forall i=1,\dots,n$
\begin{equation*}
y_i(t)= \sum_{j=1}^{13} c_{ij} B^{\omega}_j (t)
\end{equation*}
with $\mathcal{T}=[0,1]$, $B^{\omega}_j (t)$ the b-spline basis system of order 4 with interior knots $\omega=(0.1,0.2,\dots,0.9)$ and  $(c_{1,1},\dots,c_{1,13})^T,\dots,(c_{n,1},\dots,c_{n,13})^T$ i.i.d. realizations of $C=(C_1,\dots,C_{13}) \sim N_{13}\left(\textbf{0},\Sigma \right)$ such that  $\mathrm{Var}[C_i]=0.03^2$  $\forall i \neq 7$,  $\mathrm{Var}[C_7]=0.003^2$ and $\mathrm{Cov}[C_i, C_j]=0$ for $i,j=1,\dots,13$, $i \neq j$. 
\item Scenario 3. The scenario is the previous one after contamination with outliers. Formally, $(c_{1,1},\dots,c_{1,13})^T,\dots,(c_{n,1},\dots,c_{n,13})^T$ are i.i.d. realizations of a vector random variable whose probability density function is a Gaussian mixture density with weights ($1-\beta,\beta$), shared mean vector \textbf{0}, the covariance matrix defined as in Scenario 2 for the first group and such that $\mathrm{Var}[C_7]=0.3^2$ instead of $\mathrm{Var}[C_7]=0.003^2$ for the second group.
\end{itemize}
A graphical representation of a replication for each scenario with $n=18$ is provided in Figure \ref{fig:scenarios}.
The Conformal approach presented in Section \ref{sec:conformal_prediction_bands} is evaluated in the non-smoothed framework and considering three different modulation functions: $s^0$,  the normalized standard deviation function $s^{\sigma}_{\mathcal{I}_1}$ as natural representative of functions that capture data variability,  and $\bar{s}_{\mathcal{I}_1}$. Since the focus of the work is not on the construction of sophisticated point predictors $g_{\mathcal{I}_1}$ but rather on the construction of valid prediction bands around any point predictor $g_{\mathcal{I}_1}$, we hereby simply set $g_{\mathcal{I}_1}(t)=\bar{y}_{\mathcal{I}_1}(t)$.

The performance of our approach is compared to four alternative methods. These are: \textit{Naive} method, which outputs prediction bands defined as $\{y \in \mathcal{Y(T)}: y(t) \in [ q_{\frac{\alpha}{2}} \left(t \right), q_{1-\frac{\alpha}{2}} \left(t\right)]$  $\forall t \in \mathcal{T} \}$ 
with $q_{\alpha} \left(t\right)$ empirical quantile of order $\alpha$ for  $\left(y_1(t),\dots,y_n(t) \right)$. Such approach represents a very naive solution to the prediction task we are considering and we expect it to suffer greatly from undercoverage; \textit{BD} and \textit{MBD} methods, which output the sample $(1-\alpha)$  central region induced by the band depth (BD) and the modified band depth (MBD) respectively \citep{sun_functional_2011}; \textit{Boot.} method, which outputs the band based on 2500 bootstrap samples, as proposed by \citet{degras_simultaneous_2011}.
We consider $\alpha=0.1$, $\beta=0.06$ and three different sample sizes: $n=18$, $n=198$, $n=1998$. In order not to overcomplicate the simulation study, the ratio $\rho=l/n$ is kept fixed and equal to 0.5 as commonly suggested in the Conformal literature. A deeper investigation about the possible effect of the ratio $\rho=l/n$ on efficiency  - even though possibly interesting - is out of the scope of this work.
The atypical values of $n$ in the simulations have been simply chosen to have a miscoverage exactly equal to $\alpha$ (indeed in these cases $\lfloor (l+1)\alpha \rfloor/(l+1)=\alpha$) and consequently making the simulation results easier to read. Similar results  would have been attained with rounded values of $n$ (e.g. $n=20$, $n=200$, $n=2000$) by evaluating the empirical miscoverage considering the theoretical one: $\lfloor (l+1)\alpha \rfloor/(l+1)$ (see Theorem \ref{th_conservativeness_up_low_bound}). 
The simulations are achieved by using the R Programming Language \citep{R_cit} and the computation of the band depth and the modified band depth by \textit{roahd} package \citep{R_roahd}. Finally, every combination of scenario and sample size is evaluated considering $N=500$ replications.

\subsection{Coverage}
\label{sec:sim_emp_cov}

In this section we focus on the sample mean and the standard deviation of the empirical conditional coverage provided by the prediction bands generated by each method for each combination of sample size and scenario (see Table \ref{tab:validity}). 

\begin{table}[t]
\centering
\fbox{%
\begin{tabular}{ccccc c c c c}
\multicolumn{2}{c}{}&\multicolumn{3}{c}{\textbf{Conformal Method}}&\multicolumn{4}{c}{\textbf{Alternative Methods}}\\
\hline
\multicolumn{2}{c}{}&$s^0$&$s^{\sigma}_{\mathcal{I}_1}$&\multicolumn{1}{c}{$\bar{s}_{\mathcal{I}_1}$}&Naive&MBD&BD& Boot.\\ \hline
$n=18$&Sc. 1&\cellcolor{light-gray} 0.902&\cellcolor{light-gray}0.900 &\cellcolor{light-gray}0.900 &0.409 &0.504 &0.547 &0.875\\ 
&&\cellcolor{light-gray}(0.088)&\cellcolor{light-gray}(0.085) &\cellcolor{light-gray}(0.087) &(0.092) &(0.109) &(0.111) &(0.064) \\ 
&Sc. 2&\cellcolor{light-gray}0.901&0.910 &\cellcolor{light-gray}0.909 &0.048 &0.123 &0.145 &0.922\\ 
&&\cellcolor{light-gray}(0.089)&(0.081) &\cellcolor{light-gray}(0.083) &(0.021) &(0.044) &(0.051) &(0.042) \\ 
&Sc. 3&\cellcolor{light-gray}0.904&\cellcolor{light-gray}0.904&\cellcolor{light-gray}0.907&0.049 &0.124 &0.148 &0.932\\ 
&&\cellcolor{light-gray}(0.084)&\cellcolor{light-gray}(0.089) &\cellcolor{light-gray}(0.085) &(0.023) &(0.049) &(0.055) &(0.061) \\  \hline
$n=198$&Sc. 1&\cellcolor{light-gray} 0.901&\cellcolor{light-gray}0.902 &\cellcolor{light-gray}0.901 &0.625 &0.861 &\cellcolor{light-gray}0.900 &0.865\\ 
&&\cellcolor{light-gray}(0.029)&\cellcolor{light-gray}(0.030) &\cellcolor{light-gray}(0.031) &(0.031) &(0.028) &\cellcolor{light-gray}(0.028) &(0.019) \\ 
&Sc. 2&\cellcolor{light-gray}0.901&\cellcolor{light-gray}0.899 &\cellcolor{light-gray}0.900 & 0.189 &0.733 &0.788 &0.897\\ 
&&\cellcolor{light-gray}(0.029)&\cellcolor{light-gray}(0.031) &\cellcolor{light-gray}(0.029) &(0.019) &(0.036) &(0.032) &(0.015) \\ 
&Sc. 3&\cellcolor{light-gray}0.897&\cellcolor{light-gray}0.900 &\cellcolor{light-gray}0.899 &0.197 &0.742 &0.798 &0.892\\ 
&&\cellcolor{light-gray}(0.031)&\cellcolor{light-gray}(0.030) &\cellcolor{light-gray}(0.031) &(0.020) &(0.034) &(0.030) &(0.020) \\  \hline
$n=1998$&Sc. 1&\cellcolor{light-gray}0.900&\cellcolor{light-gray}0.899 &\cellcolor{light-gray}0.900 & 0.666 &0.942 &0.918 &0.866\\ 
&&\cellcolor{light-gray}(0.010)&\cellcolor{light-gray}(0.010) &\cellcolor{light-gray}(0.010) &(0.011) &(0.006) &(0.008) &(0.008) \\ 
&Sc. 2&\cellcolor{light-gray}0.900&\cellcolor{light-gray}0.900 &\cellcolor{light-gray}0.899 &0.233 &0.958 &0.971 &\cellcolor{light-gray}0.899\\ 
&&\cellcolor{light-gray}(0.009)&\cellcolor{light-gray}(0.010) &\cellcolor{light-gray}(0.010) &(0.007) &(0.006) &(0.005) &\cellcolor{light-gray}(0.008) \\ 
&Sc. 3&\cellcolor{light-gray}0.900&\cellcolor{light-gray}0.899 &\cellcolor{light-gray}0.900 &0.240 &0.959 &0.973 &0.884\\ 
&&\cellcolor{light-gray}(0.010)&\cellcolor{light-gray}(0.010) &\cellcolor{light-gray}(0.010) &(0.008) &(0.006) &(0.005) &(0.007) \\  
\end{tabular} }
\caption{For each combination of sample size and scenario, the first line shows the sample mean of the empirical conditional coverage, the second line the sample standard deviation in brackets. A combination of mean and st. deviation is gray-colored if the corresponding 99\% confidence t-interval for the (unconditional) coverage includes value $1-\alpha$.}
\label{tab:validity}
\end{table}
Specifically, the  empirical conditional coverage of a given prediction band (i.e. the empirical coverage obtained conditioning on the prediction band obtained by the observed data) is computed as the fraction of times that 10,000 new functions - independent from and identically distributed to the original sample - belong to such prediction band. The purpose of this scheme is twofold: first of all, by averaging the $N=500$  empirical conditional coverages obtained for  each combination of scenario and sample size it is possible to obtain the empirical coverage, which is an estimate of  the (unconditional) coverage. Secondly, this scheme allows to evaluate the variability of the conditional coverage when the observed sample varies, a particularly useful indication in real applications. In order to facilitate the visualization of the results and to allow inferential conclusions, a specific combination of sample mean and standard deviation is gray-colored in Table \ref{tab:validity} if the corresponding 99\% confidence t-interval for the (unconditional) coverage includes 0.90, i.e. the value $1-\alpha$. 

The simulation study fully confirms the theoretical property concerning the validity of split conformal prediction sets with 53 out of the 54 99\%-confidence intervals associated to conformal bands including the nominal value $1-\alpha$. The evidence provided is particularly appealing since the desired coverage is guaranteed also when a very small sample size ($n=18$) is considered, a framework in which such property is traditionally hard to obtain. Vice versa, in almost all cases the alternative methods do not ensure the desired coverage with some estimates  dramatically far from $1-\alpha$, especially for small sample sizes (i.e., $n=18$). In view of this, in Section \ref{sec:sim_eff_study} only the efficiency of the Conformal methods is evaluated and compared.

\subsection{Efficiency}
\label{sec:sim_eff_study}

In this section the sample mean and the standard deviation of the size defined as in \eqref{eq:area} of the prediction bands computed in the previous section are evaluated for each combination of modulation function, sample size and scenario (see Table \ref{tab:size}).  

\begin{table}
\centering
\fbox{%
\begin{tabular}{ccccc c c c c}
\multicolumn{2}{c}{}&\multicolumn{2}{c}{$s^0$}&\multicolumn{2}{c}{$s^{\sigma}_{\mathcal{I}_1}$}&\multicolumn{2}{c}{$\bar{s}_{\mathcal{I}_1}$}\\
\cline{3-4}
\cline{5-6}
\cline{7-8}
&&$Mean$&$st.dev$&$Mean$&$st.dev$&$Mean$&$st.dev$\\ \hline
\textbf{$n=18$}&Sc. 1&\cellcolor{light-gray}8.113&(2.044) &10.088 &(3.618) & 11.638 &(4.309) \\
&Sc. 2&\cellcolor{light-gray}0.142&(0.025) &0.165 &(0.041) &0.185 &(0.049)\\  
&Sc. 3&\cellcolor{light-gray}0.246&(0.192) &0.448 &(0.550) &0.505 &(0.633)\\   
\textbf{$n=198$}&Sc. 1&\cellcolor{light-gray}7.175&(0.560) &7.295 &(0.608) &7.556 &(0.647)\\  
&Sc. 2&0.127&(0.006) &\cellcolor{light-gray}0.109 &(0.005) &0.120 &(0.006)\\  
&Sc. 3&0.139&(0.013) &0.139 &(0.013) &\cellcolor{light-gray}0.137 &(0.020)\\   
\textbf{$n=1998$}&Sc. 1&\cellcolor{light-gray}7.059&(0.179) &7.065 &(0.176) &7.128 &(0.184) \\  
&Sc. 2&0.125&(0.002) &\cellcolor{light-gray} 0.106&(0.001) &0.117 &(0.002) \\  
&Sc. 3&0.136&(0.003) & 0.137 &(0.004) &\cellcolor{light-gray}0.131 &(0.003)\\     
\end{tabular} }
\caption{Size of the prediction bands. For each row, the lowest value of the sample mean is gray-colored.}
\label{tab:size}
\end{table}

First of all, it is noticeable that when $n=18$ the absence of modulation (i.e. $s^0$) seems to provide smaller prediction bands than those induced by $s^{\sigma}_{\mathcal{I}_1}$ and $\bar{s}_{\mathcal{I}_1}$, conceivably because the extremely low number of functions belonging to the training set  ($m=9$) leads to an unstable and possibly misleading modulation function supporting the statistical intuition that for small sample sizes simpler modulation functions should be preferred. 

More deeply, focusing now on each scenario separately and considering the remaining sample sizes, Scenario 1 represents a framework in which a constant width prediction band is the ideal candidate since the horizontal shift due to the random variable $U$ induces constant variance along the domain. As a consequence,  the pointwise evaluations $Y(t)$ are equally distributed $\forall t \in \mathcal{T}$ and so one is justified in expecting $s^{\sigma}_{\mathcal{I}_1}$ and $\bar{s}_{\mathcal{I}_1}$ to be of no practical use.
The results confirm this conjecture, but the differences between the three  modulation functions seems to decrease as the sample size grows (see, for example, the difference between $s^0$ and $\bar{s}_{\mathcal{I}_1}$ when $n$ increases from 198 to 1998).

Scenario 2 represents a completely different setting, in which a modulation process is appropriate since the curves highlight a reduction of variability in the central part of the domain. As expected, $s^0$ induces larger predictions bands (on average) than those obtained by $s^{\sigma}_{\mathcal{I}_1}$ and $\bar{s}_{\mathcal{I}_1}$ and it forces the band to be unnecessary large around $t=0.5$. On the other hand, the other two modulation functions (especially $s^{\sigma}_{\mathcal{I}_1}$) provide a better performance since they allow the band width to be adapted according to the behavior of data over $\mathcal{T}$.

\begin{figure}[!t]
\begin{center}
\includegraphics[width=14.75cm,height=9.680cm]{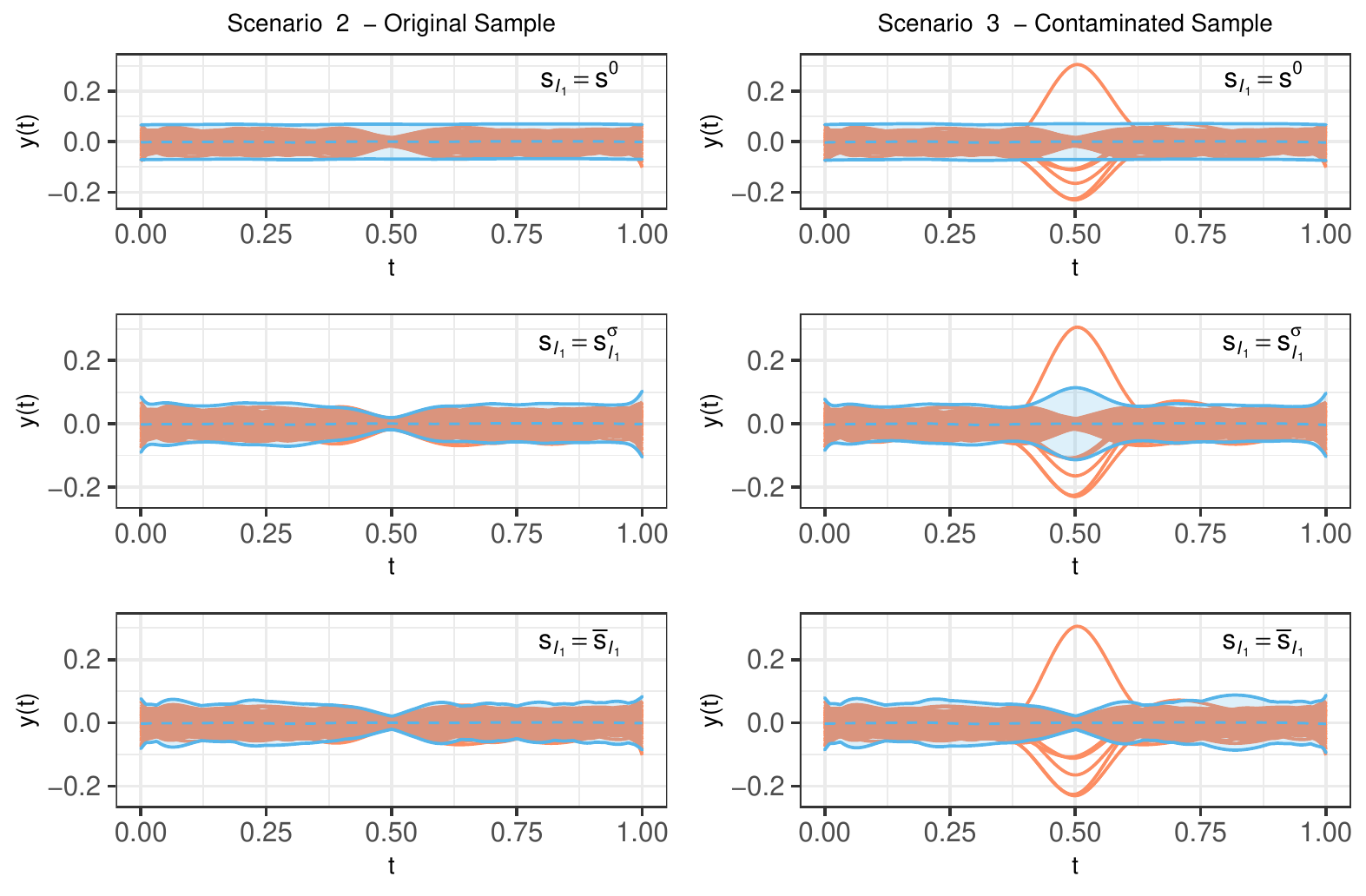}
\end{center}
\caption{The prediction bands obtained considering a combination of  modulation functions ($s^0$ at the top, $s^{\sigma}_{\mathcal{I}_1}$ in the middle, $\bar{s}_{\mathcal{I}_1}$ at the bottom) and sample (the original one on the left, the contaminated one on the right). In all cases, the dashed line represents $g_{\mathcal{I}_1}$.}
\label{fig:outliers}
\end{figure}

Scenario 3 is obtained by contaminating Scenario 2 with outliers. Table  \ref{tab:size} suggests that $\bar{s}_{\mathcal{I}_1}$ outperforms both $s^0$ and - unlike Scenario 2 - also $s^{\sigma}_{\mathcal{I}_1}$. In order to clarify this evidence, let us consider a sample $y_1,\dots,y_{198}$ generated as in Scenario 2 that, after being created, is exposed to a contamination process in which each function $y_i, i=1,\dots,198,$ becomes an outlier as described in Scenario 3 with probability $\beta=0.06$. Figure \ref{fig:outliers}  
shows examples of prediction bands induced by the three modulation functions ($s^0$ at the top, $s^{\sigma}_{\mathcal{I}_1}$ in the middle, $\bar{s}_{\mathcal{I}_1}$ at the bottom) obtained by considering the original sample (on the left) and the contaminated one (on the right). Moving from Scenario 2 to Scenario 3 and focusing on  $s^{\sigma}_{\mathcal{I}_1}$, it is possible to notice that the increased variability in the central part of the domain due to the contamination process involves an increase in the band width around $t=0.5$. This behavior, although not surprising, is counterproductive since the purpose of the method is to create prediction bands with coverage at the level $1-\alpha=0.9$ and in this specific case $\sim 94 \%$ of the functions tends to be highly concentrated around $g_{\mathcal{I}_1}$ in the central part of the domain, and not overdispersed. By contrast, $\bar{s}_{\mathcal{I}_1}$ by construction removes the most extreme (in terms of measure \eqref{eq:original_NCM}) functions and properly modulates data on the basis of the non-extreme functions keeping the band shape unchanged. From a methodological point of view, this is due to the dependency of $\bar{s}_{\mathcal{I}_1}$ on $\alpha$ which allows only a portion of the training set - chosen according to the specific level $1-\alpha$ - to be taken into account  and the trend of the ``misleading'' functions to be completely ignored.
Overall, the evidence provided by this example - together with the results provided by Table \ref{tab:size} - suggests that  $s^0$ is not affected by the contamination process (pro) but does not modulate (con), $s^{\sigma}_{\mathcal{I}_1}$ modulates (pro) but overreacts to the contamination process (con), whereas $\bar{s}_{\mathcal{I}_1}$ is able to simultaneously modulate (pro) and manage the contamination process (pro).

In short, the three scenarios seem to highlight that $s^0$ is an outstanding candidate when the sample size is very small, whereas a modulation process is useful in the very common case in which the variability over $\mathcal{T}$ varies and the sample size is either moderate or large. Specifically,
$\bar{s}_{\mathcal{I}_1}$ provides encouraging results in some complex scenarios as it focuses on the specific behavior of the central (according to the level $1-\alpha$) portion of data.

\section{Application}
\label{sec:application}

In order to show the wide generality of our approach, in this section we apply our Conformal approach to a well known data set in the FDA community (i.e., the Berkeley Growth Study data set \citep{tuddenham1954physical}) that is characterized by features that cannot be trivially framed in a standard probabilistic parametric model, i.e.: heteroscedasticity along the functional domain, phase misalignment, presence of outlier curves, and positivity constraint. The specific data set contains in detail the heights (in cm) of 54 female and 39 male children measured quarterly from 1 to 2 years, annually from 2 to 8 years and biannually from 8 to 18 years. We focus on the first derivative of the growth curves, which are estimated in a standard fashion by R function \textit{smooth.monotone} of \textit{fda} package \citep{fda_package} implementing monotonic cubic regression splines \citep[][chap. 6]{ramsay_functional_2005}.
Specifically, the prediction bands here reported refer to the growth velocity curves between 4 and 18 years for girls and boys separately comparing, in the Non-Smoothed Conformal framework, the three modulation functions analyzed in Section \ref{sec:simulation_study} and with  $g_{\mathcal{I}_1}$ being simply for each group the corresponding functional sample mean, $\alpha=0.5$, $m=27$ for girls, $m=20$ for boys. 

\begin{figure}[!t]
\begin{center}
\includegraphics[width=14.75cm,height=7.836cm]{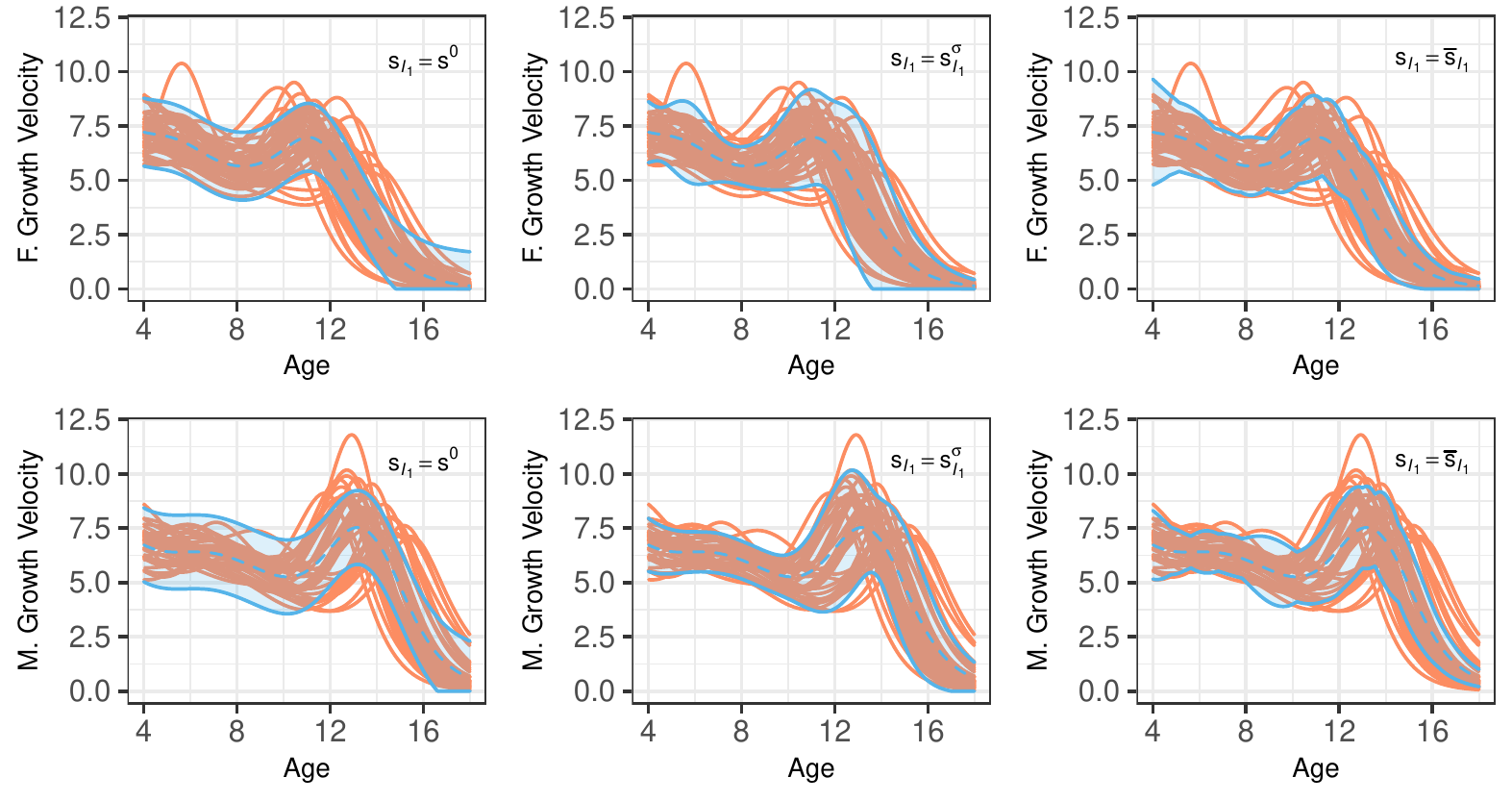}
\end{center}
\caption{Berkeley Growth Study data: each panel shows the prediction band obtained considering a different modulation function ($s^0$ on the left, $s^{\sigma}_{\mathcal{I}_1}$ in the middle, $\bar{s}_{\mathcal{I}_1}$ on the right).  In all cases, the dashed line represents $g_{\mathcal{I}_1}$. Predictions for girls at the top and predictions for boys at the bottom.}
\label{fig:app_fem_mal}
\end{figure}

The prediction bands are shown in Figure \ref{fig:app_fem_mal}.  
Note that since the application at hand does not allow the functions to be negative in any subset of the domain, the prediction bands can be (and are indeed) truncated to 0 without decreasing their coverage. The possibility of removing from the prediction bands regions which are known  - from the domain knowledge - to have null probability, without affecting the coverage, is a desirable implication derived from using a fully nonparametric approach to prediction since this takes away the burden of an explicit and possibly non-trivial modeling of lower and/or upper bound constraints.

Focusing on Figure \ref{fig:app_fem_mal}, the graphical representation of the prediction bands highlights the well-known different growth path between girls and boys, in which the latter group typically starts to grow later but achieves higher growth velocities. In terms of the role of modulation functions, their impact on female growth velocity prediction seems to be less than the one on the male bands. From a prediction point of view, girls' curves represent a simpler scenario in which the variance is lower along the domain, while boys' curves represent a more tricky scenario with strong heteroscedasticity of the functions over $\mathcal{T}$ (due to the joint presence of misalignment of data and a very localized high peak around 13 years of age). As expected from these considerations, the prediction bands for a new girl's velocity curve obtained using the different modulation functions are relatively similar, with the prediction band associated to  $\bar{s}_{\mathcal{I}_1}$ being aslightly narrower due to the presence of outliers. Instead focusing on boys' curves, the strong heteroscedasticity forces the prediction band induced by $s^0$ to be uselessly large in some parts of the domain, whereas in general the prediction band induced by $s^{\sigma}_{\mathcal{I}_1}$ seems to be smoother than that induced by  $\bar{s}_{\mathcal{I}_1}$, whose ``bumps'' are caused by the specific  modulation function used. Both for boys and girls $\bar{s}_{\mathcal{I}_1}$ outputs the smallest prediction band, as shown in Table \ref{tab:size_application} where the quantity $\mathcal{Q}(\cdot)/|\mathcal{T}|$ is reported. 

\begin{table}
\caption{\label{tab:size_application}Berkeley Growth Study data: average width of the prediction bands.}
\centering
\fbox{%
\begin{tabular}{c|ccc}
&$s^0$&$s^{\sigma}_{\mathcal{I}_1}$&$\bar{s}_{\mathcal{I}_1}$\\ \hline
Females &2.904&3.244&2.811\\ 
Males &3.334&3.107&2.690\\ 
\end{tabular} }
\end{table}

\begin{figure}[]
\begin{center}
\includegraphics[width=6.27cm,height=5cm]{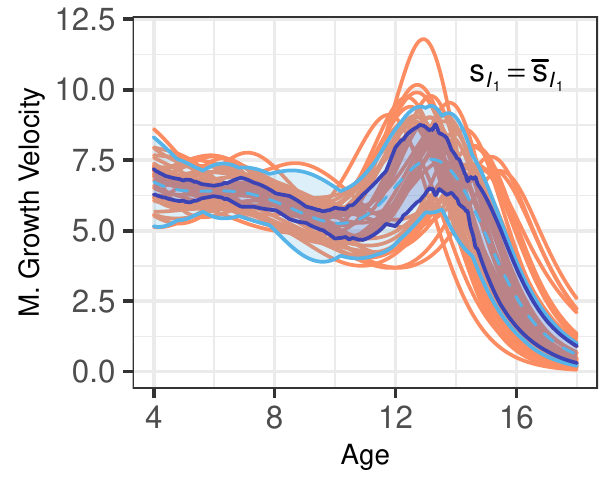}
\end{center}
\caption{Berkeley Growth Study data: the prediction band represented at the bottom right of Figure \ref{fig:app_fem_mal} (light blue) and the correponding pointwise conformal prediction band (dark blue).}
\label{fig:app_pointwise}
\end{figure}

Some useful information can be also provided by the comparison between the proposed approach and its pointwise counterpart, in which the prediction band is constructed by applying a coherent univariate Conformal approach at each point $t$ separately. Indeed, by construction the former creates prediction bands larger or equal than those obtained by the latter, but on the other hand it guarantees simultaneous (and not pointwise) validity and of course it interprets a function as a whole, a key aspect in the functional context. In order to clarify this concept, let us consider Figure \ref{fig:app_pointwise}, 
in which the pointwise prediction band (dark blue) is overlaid to the bottom-right panel of Figure \ref{fig:app_fem_mal}. As expected, the pointwise prediction band is simply modulated by the local variability of the 50\% central curves. Differently, the prediction bands here proposed instead take also into consideration the behavior of the functions along the domain $\mathcal{T}$ with the effect of generating narrower or wider bands also in presence of similar local variabilities and so not just obtaining a simple expansion of the pointwise prediction band.

\section{Conclusion}
\label{sec:discussion}

The creation of prediction sets for functional data is still an open problem of paramount importance in statistical methodology research. In order to define and compute them, the great majority of methods currently presented in the literature rely on non-provable distributional assumption, dimension reduction techniques and/or asymptotic arguments. On the contrary, the approach proposed in this article represents an innovative proposal in this field: indeed, the Conformal framework ensures that finite-sample either valid or exact prediction sets are obtained under minimal distributional assumptions, whereas the specific family of nonconformity measures introduced guarantees - besides prediction sets that are  bands - also a fast, scalable and closed-form solution. Moreover, despite the fact that our approach works regardless the specific choice of  $s_{\mathcal{I}_1}$ (which can be chosen, for example, a priori), we proposed a specific data-driven modulation function, namely $ \bar{s}_{\mathcal{I}_1}$, which leads to prediction bands asymptotically no less efficient than those obtained by not modulating. The focus of this article was on i.i.d. data, but we envision an extension of the procedure to regression and classification problems.

Our procedure is able to achieve encouraging results and could represent a promising starting point for future developments, but at least two aspects, among others, should be carefully investigated. First of all, the division of data into the training and calibration sets induces an intrinsic element of randomness  into the method and, although this phenomenon is well known in the Conformal literature, a quantification of the effect of the split process - and also of the values $m$ and $l$ - on the procedure has not yet been properly analyzed. Secondly, the prediction sets proposed in this article are purposely shaped as functional bands. This geometrical characterization in most applicative scenarios can be considered well suited. Nevertheless, one can think at more complicated scenarios (e.g., functional mixtures) where prediction set made of multiple bands could be considered more suited from an application point of view. This possible extension will be the object of future work. 


\section*{Supplementary material}

\subsection*{A.1 Proofs of Section \ref{sec:conformal_approach}}

\textbf{Proof of Theorem \ref{th_conservativeness_up_low_bound}}. 

Since $\mathcal{C}_{n,1-\alpha}:= \left\{y \in \mathcal{Y(T)}: \delta_{y}>\alpha \right\}$, then $\mathcal{C}_{n,1-\alpha}:= \left\{y \in \mathcal{Y(T)}: (l+1) \delta_{y}> (l+1)\alpha \right\}$. Under the hypothesis of the theorem, $ (l+1) \delta_{Y} \sim U\{ 1,2,\dots,l+1 \}$ holds. As a consequence:
\begin{align*}
\mathbb{P} \left( Y_{n+1} \in \mathcal{C}_{n,1-\alpha} \right) &=\mathbb{P} \left(  (l+1) \delta_{Y}> (l+1)\alpha \right) \\
&=1-\mathbb{P} \left(  (l+1) \delta_{Y} \leq (l+1)\alpha \right) \\
&=1-\frac{\lfloor (l+1)\alpha \rfloor}{l+1}.
\end{align*}
In addition, since 
\begin{equation*}
\frac{\lfloor (l+1)\alpha \rfloor}{l+1} \leq \frac{ (l+1)\alpha}{l+1}= \alpha
\end{equation*}
then  $\mathbb{P} \left( Y_{n+1} \in \mathcal{C}_{n,1-\alpha} \right) \geq 1 - \alpha$, i.e. $\mathcal{C}_{n,1-\alpha}$ is valid. Finally, since 
\begin{equation*}
\frac{\lfloor (l+1)\alpha \rfloor}{l+1} > \frac{ (l+1)\alpha -1}{l+1}= \alpha - \frac{1}{l+1}
\end{equation*}
then $\mathbb{P} \left( Y_{n+1} \in \mathcal{C}_{n,1-\alpha} \right) < 1 - \alpha + \frac{1}{l+1}$.

\textbf{Proof that smoothed split conformal prediction sets are exact.}

Let us consider the hypothesis of Theorem \ref{th_conservativeness_up_low_bound}. Let us notice that
\begin{align*}
\delta_{y,\tau_{n+1}} :=&  \frac{\left|\left\{j \in  \mathcal{I}_2: R_{j} > R_{n+1}\right\}\right| + \tau_{n+1} \left|\left\{j \in  \mathcal{I}_2 \cup \{n+1\}: R_{j} = R_{n+1}\right\}\right|}{l+1} \\
=&\frac{\tau_{n+1}}{l+1} +  \frac{\left|\left\{j \in  \mathcal{I}_2: R_{j} \geq R_{n+1}\right\}\right|}{l+1}.
\end{align*}
Under the hypothesis of Theorem \ref{th_conservativeness_up_low_bound}, $\left|\left\{j \in  \mathcal{I}_2: R_{j} \geq R_{n+1}\right\}\right|  \sim U\{ 0,1,\dots,l \}$ holds. As a consequence:
\begin{align*}
\mathbb{P} \left( Y_{n+1} \in \mathcal{C}_{n,1-\alpha,\tau_{n+1}} | \tau_{n+1} \right) &=\mathbb{P} \left( \delta_{Y,\tau_{n+1}}> \alpha | \tau_{n+1} \right) \\
&=\mathbb{P} \left(  \left|\left\{j \in  \mathcal{I}_2: R_{j} \geq R_{n+1}\right\}\right|   > (l+1)\alpha - \tau_{n+1} | \tau_{n+1} \right) \\
&=1-\mathbb{P} \left(  \left|\left\{j \in  \mathcal{I}_2: R_{j} \geq R_{n+1}\right\}\right|   \leq (l+1)\alpha - \tau_{n+1} | \tau_{n+1} \right) \\
&=1-\frac{\lfloor (l+1)\alpha - \tau_{n+1} \rfloor +1}{l+1}.
\end{align*}
Let us call $f(\tau_{n+1})=1 \cdot \mathds{1} \{\tau_{n+1} \in [0,1] \}$. Then
\begin{align*}
\mathbb{P} \left( Y_{n+1} \in \mathcal{C}_{n,1-\alpha,\tau_{n+1}} \right) =& \int_{0}^1 \mathbb{P} \left( Y_{n+1} \in \mathcal{C}_{n,1-\alpha,\tau_{n+1}} | \tau_{n+1} \right) f(\tau_{n+1}) d\tau_{n+1} \\
=&1 - \\
&\Bigg( \int_{0}^{(l+1)\alpha - \lfloor (l+1)\alpha \rfloor} \frac{\lfloor (l+1)\alpha - \tau_{n+1} \rfloor +1}{l+1} d\tau_{n+1} + \\
&\int_{(l+1)\alpha - \lfloor (l+1)\alpha \rfloor}^1 \frac{\lfloor (l+1)\alpha - \tau_{n+1} \rfloor +1}{l+1}  d\tau_{n+1} \Bigg). \\
\end{align*}

Let us consider $ \int_{0}^{(l+1)\alpha - \lfloor (l+1)\alpha \rfloor} \frac{\lfloor (l+1)\alpha - \tau_{n+1} \rfloor +1}{l+1} d\tau_{n+1}$. Since if $\tau_{n+1} \leq (l+1)\alpha - \lfloor (l+1)\alpha \rfloor $ then $\lfloor (l+1)\alpha - \tau_{n+1} \rfloor= \lfloor (l+1)\alpha \rfloor$, we can notice that
\begin{align*}
&\int_{0}^{(l+1)\alpha - \lfloor (l+1)\alpha \rfloor} \frac{\lfloor (l+1)\alpha - \tau_{n+1} \rfloor +1}{l+1} d\tau_{n+1}\\
=&\int_{0}^{(l+1)\alpha - \lfloor (l+1)\alpha \rfloor} \frac{\lfloor (l+1)\alpha\rfloor +1}{l+1} d\tau_{n+1}\\
=&\frac{ \lfloor (l+1)\alpha \rfloor +1}{l+1} \cdot \left( (l+1)\alpha - \lfloor (l+1)\alpha \rfloor \right). \\
\end{align*}

Let us consider $\int_{(l+1)\alpha - \lfloor (l+1)\alpha \rfloor}^1 \frac{\lfloor (l+1)\alpha - \tau_{n+1} \rfloor +1}{l+1}  d\tau_{n+1}$. Since if $\tau_{n+1} > (l+1)\alpha - \lfloor (l+1)\alpha \rfloor $ then $\lfloor (l+1)\alpha - \tau_{n+1} \rfloor= \lfloor (l+1)\alpha \rfloor -1$, we can notice that
\begin{align*}
&\int_{(l+1)\alpha - \lfloor (l+1)\alpha \rfloor}^1 \frac{\lfloor (l+1)\alpha - \tau_{n+1} \rfloor +1}{l+1}  d\tau_{n+1} \\
=&\int_{(l+1)\alpha - \lfloor (l+1)\alpha \rfloor}^1 \frac{\lfloor (l+1)\alpha \rfloor}{l+1}  d\tau_{n+1} \\
=&\frac{ \lfloor (l+1)\alpha \rfloor}{l+1} \cdot \left( 1- \left((l+1)\alpha - \lfloor (l+1)\alpha \rfloor \right) \right). \\
\end{align*}
Then
\begin{align*}
&\mathbb{P} \left( Y_{n+1} \in \mathcal{C}_{n,1-\alpha,\tau_{n+1}} \right) \\
=&1 - \\
&\Bigg( \frac{ \lfloor (l+1)\alpha \rfloor +1}{l+1} \cdot \left( (l+1)\alpha - \lfloor (l+1)\alpha \rfloor \right) + \\
&\frac{ \lfloor (l+1)\alpha \rfloor}{l+1} \cdot \left( 1- \left((l+1)\alpha - \lfloor (l+1)\alpha \rfloor \right) \right) \Bigg)\\
=&1-\alpha.
\end{align*}

\subsection*{A.2 Proofs of Section \ref{sec:supremum_metric}}

\textbf{Proof that the concatenation of pointwise prediction intervals leads to a prediction band that is a subset of the simultaneous prediction band (\ref{eq:pred_set_original})}. 

Let $\mathcal{U}_{n, 1-\alpha}$ be the pointwise prediction set. Let us define  $\tilde{R}_j(t):=\left| y_{j}(t)-g_{\mathcal{I}_1}(t)\right|$ $\forall t \in \mathcal{T}, j \in \mathcal{I}_2$, $\tilde{R}_{n+1}(t):=\left| y(t)-g_{\mathcal{I}_1}(t)\right|$ for a given $y \in \mathcal{Y(T)}$ and   $\tilde{k}(t)$ the $\lceil (l+1)(1-\alpha) \rceil$th smallest value in the set $\{ \tilde{R}_h(t): h \in \mathcal{I}_2 \}$.
By construction $R_j=\sup_{t \in \mathcal{T}} \tilde{R}_j(t)$, and so $R_j \geq \tilde{R}_j(t)$ $\forall t \in \mathcal{T}, j \in \mathcal{I}_2$ and then $k \geq \tilde{k}(t)$ $\forall t \in \mathcal{T}$.
Let us consider $y \in \mathcal{U}_{n, 1-\alpha}$, i.e. $y(t) \in [g_{\mathcal{I}_1}(t)-\tilde{k}(t), g_{\mathcal{I}_1}(t)+\tilde{k}(t)] \quad \forall t \in \mathcal{T} $. Since $k \geq \tilde{k}(t)$, also $y(t) \in [g_{\mathcal{I}_1}(t)-k, g_{\mathcal{I}_1}(t)+k] \quad \forall t \in \mathcal{T}$, i.e. $y \in \mathcal{C}_{n, 1-\alpha}$. 

Since the converse is not necessarily true (in the sense that  $y \in \mathcal{C}_{n, 1-\alpha}$ does not imply $y \in \mathcal{U}_{n, 1-\alpha}$), we conclude that $\mathcal{U}_{n, 1-\alpha} \subseteq \mathcal{C}_{n, 1-\alpha}$.

\subsection*{A.3 Proofs of Section \ref{sec:modulation}}

\textbf{Proof of the prediction set  induced by the nonconformity measure  $A( \{y_h: h \in  \mathcal{I}_1 \},y)= \sup_{t \in \mathcal{T}} \left| \frac{y(t)-g_{\mathcal{I}_1}(t)}{s_{\mathcal{I}_1}(t)}\right|$. }

For a given $y \in \mathcal{Y(T)}$, let us define
\begin{equation*}
\delta^{s}_y :=  \frac{\left|\left\{j \in  \mathcal{I}_2 \cup \{n+1\} : R^{s}_{j} \geq R^{s}_{n+1}\right\}\right|}{l+1}.
\end{equation*}
The split conformal prediction set is defined as $\mathcal{C}_{n, 1-\alpha}^{s}:= \left\{y \in \mathcal{Y(T)}: \delta^{s}_{y}>\alpha \right\}$. As a consequence, 
$y \in \mathcal{C}_{n, 1-\alpha}^{s} \iff R^{s}_{n+1} \leq k^s$, with  $k^s$ the $\lceil (l+1)(1-\alpha) \rceil$th smallest value in the set $\{ R^{s}_h: h \in \mathcal{I}_2 \}$.
Then:
\begin{align*}
\sup_{t \in \mathcal{T}} &\left| \frac{y(t)-g_{\mathcal{I}_1}(t)}{s_{\mathcal{I}_1}(t)}\right| \leq k^s \\
\iff &\left| \frac{y(t)-g_{\mathcal{I}_1}(t)}{s_{\mathcal{I}_1}(t)}\right| \leq k^s \quad \forall t \in \mathcal{T}  \\
\iff & y(t) \in [g_{\mathcal{I}_1}(t)-k^s s_{\mathcal{I}_1}(t) , g_{\mathcal{I}_1}(t)+k^s s_{\mathcal{I}_1}(t)] \quad \forall t \in \mathcal{T}. 
\end{align*}

Therefore, the  split conformal prediction set is
\begin{equation*}
\mathcal{C}^s_{n, 1-\alpha}:= \left\{y \in \mathcal{Y(T)}: y(t) \in [g_{\mathcal{I}_1}(t)-k^s s_{\mathcal{I}_1}(t) , g_{\mathcal{I}_1}(t)+k^s s_{\mathcal{I}_1}(t)] \quad \forall t \in \mathcal{T}  \right\}.
\label{eq:pred_set_modulated}
\end{equation*}

\textbf{Proof of Remark \ref{remark_integral_equal_1}}. 

Let us define $\mathcal{C}_{n, 1-\alpha}^{\lambda \cdot s}$ the prediction set obtained by considering the modulation function $\lambda \cdot s_{\mathcal{I}_1}$. The nonconformity scores are
\begin{align*}
R^{\lambda \cdot s}_j=& \sup_{t \in \mathcal{T}} \left| \frac{y_{j}(t)-g_{\mathcal{I}_1}(t)}{\lambda \cdot s_{\mathcal{I}_1}(t)}\right|=\frac{1}{\lambda} R^{s}_j, \quad j \in \mathcal{I}_2\\
R^{\lambda \cdot s}_{n+1}=& \sup_{t \in \mathcal{T}} \left| \frac{y(t)-g_{\mathcal{I}_1}(t)}{\lambda \cdot s_{\mathcal{I}_1}(t)}\right|=\frac{1}{\lambda} R^{s}_{n+1}.
\end{align*}
Let us also define
\begin{equation*}
\delta^{\lambda \cdot s}_y :=  \frac{\left|\left\{j \in  \mathcal{I}_2 \cup \{n+1\} : R^{\lambda \cdot s}_{j} \geq R^{\lambda \cdot s}_{n+1}\right\}\right|}{l+1}.
\end{equation*}
The split conformal prediction set is defined as $\mathcal{C}_{n, 1-\alpha}^{\lambda \cdot s}:= \left\{y \in \mathcal{Y(T)}: \delta^{\lambda \cdot s}_{y}>\alpha \right\}$. As a consequence, 
$y \in \mathcal{C}_{n, 1-\alpha}^{\lambda \cdot s} \iff R^{\lambda \cdot s}_{n+1} \leq k^{\lambda \cdot s}$, with $k^{\lambda \cdot s}$ the $\lceil (l+1)(1-\alpha) \rceil$th smallest value in the set $\{ R^{\lambda \cdot s}_h: h \in \mathcal{I}_2 \}$. In addition, since $R^{\lambda \cdot s}_j= R^{s}_j/ \lambda$ $\forall j \in \mathcal{I}_2 $, then $k^{\lambda \cdot s}= k^s/\lambda$. Then:

\begin{align*}
&\phantom{\Rightarrow} R^{\lambda \cdot s}_{n+1} \leq k^{\lambda \cdot s} \\
&\iff \frac{1}{\lambda} R^{s}_{n+1} \leq \frac{k^s}{\lambda}  \\
&\iff R^{s}_{n+1} \leq k^s,  \\
\end{align*}

and since $y \in \mathcal{C}_{n, 1-\alpha}^{s} \iff R^{s}_{n+1} \leq k^s$, then $\mathcal{C}_{n, 1-\alpha}^{\lambda \cdot s}= \mathcal{C}_{n, 1-\alpha}^{s}$.

\textbf{Adjustment procedure of $ \bar{s}^{c}_{\mathcal{I}_1}$ and $ \bar{s}_{\mathcal{I}_1}$}

If $\max_{j \in \mathcal{H}_2} |y_{j}(t)- g_{\mathcal{I}_1}(t)|=0$ for at least one value $t$ but the condition $\int_{\mathcal{T}} \max_{j \in \mathcal{H}_2} |y_{j}(t)- g_{\mathcal{I}_1}(t)| dt \neq 0$ still holds, in order to ensure that $\bar{s}^{c}_{\mathcal{I}_1}(t)>0$ $\forall t \in \mathcal{T}$ it is sufficient to add an arbitrarily (small) positive value to $\bar{s}^{c}_{\mathcal{I}_1}(t)$ $\forall t \in \mathcal{T}$ and to adjust the normalization constant accordingly. 
The pathological case in which $\int_{\mathcal{T}} \max_{j \in \mathcal{H}_2} |y_{j}(t)- g_{\mathcal{I}_1}(t)| dt = 0$ is addressed only when $y_{j}(t) = g_{\mathcal{I}_1}(t)$ $\forall j \in \mathcal{H}_2$ and almost every $t \in \mathcal{T}$ and it represents a case of no practical interest. 

Should $\exists$ $t \in \mathcal{T}$ such that $\max_{j \in \mathcal{H}_1} |y_{j}(t)- g_{\mathcal{I}_1}(t)| = 0$, the same procedure is developed.

\textbf{Proof of Theorem \ref{th:shared_convergence}}. 

Let us focus on $\bar{s}_{\mathcal{I}_1}(t)$.
Since $m/n=\theta$ with $0 < \theta < 1$, if $n \to +\infty$ then $m \to +\infty$. By definition, the scalar $\gamma$ is the empirical quantile of order $\lceil (m+1)(1-\alpha) \rceil)$ of $\{ \sup_{t \in \mathcal{T}} |y_{h}(t)-g_{\mathcal{I}_1}(t)|: h \in \mathcal{I}_1 \}$. First of all note that 
\begin{align*}
\lim_{m \to +\infty} \frac{\lceil (m+1)(1-\alpha) \rceil}{m}= \lim_{m \to +\infty} \frac{m+1 - \lfloor (m+1)\alpha \rfloor}{m}
\end{align*}
and since
\begin{equation*}
\frac{(m+1)\alpha -1}{m} \leq \frac{\lfloor(m+1)\alpha \rfloor}{m} \leq \frac{(m+1)\alpha}{m} \quad \forall m \in  \mathbb{N}
\end{equation*}
and

\begin{equation*}
\lim_{m \to +\infty} \frac{(m+1)\alpha -1}{m} = \lim_{m \to +\infty} \frac{(m+1)\alpha}{m} = \alpha
\end{equation*}

then by the squeeze theorem (also known as the sandwich theorem) we obtain that
\begin{equation*}
\lim_{m \to +\infty} \frac{\lfloor(m+1)\alpha \rfloor}{m} = \alpha
\end{equation*}
and then
\begin{align*}
\lim_{m \to +\infty} \frac{\lceil (m+1)(1-\alpha) \rceil)}{m}= 1-\alpha.
\end{align*}

As a consequence, $\gamma$ is the empirical quantile of order  $1-\alpha$ when $m \rightarrow +\infty$.

For convenience, let us define $x_i:=  \sup_{t \in \mathcal{T}} |y_{i}(t)- g_{\mathcal{I}_1}(t)|$ $\forall$ $i \in \mathcal{I}_1$. The random variables $\{X_h: h \in \mathcal{I}_1 \}$ from which $\{x_h: h \in \mathcal{I}_1 \}$ are drawn are continuous and they are asymptotically i.i.d. as $\mathrm{Var}[g_{\mathcal{I}_1}(t)] \to 0 $. The Glivenko-Cantelli theorem ensures that the empirical distribution function of these variables converges uniformly (and almost surely pointwise) to its distribution function, and then also the empirical quantiles converge in distribution (and so in probability) to the corresponding theoretical quantiles, as shown for example by \citet[][ chap. 21]{van2000asymptotic}. Specifically, empirical quantile $\gamma$ converges to $q_{1-\alpha}$, the theoretical quantile of order $1-\alpha$. As a consequence, when $m \to +\infty$:
\begin{equation*}
\mathcal{H}_1:=\{j \in \mathcal{I}_1: \sup_{t \in \mathcal{T}} |y_{j}(t)-g_{\mathcal{I}_1}(t)| \leq q_{1-\alpha}\} 
\end{equation*}
with $q_{1-\alpha}$ deterministic quantity. Let us focus on the numerator of $\bar{s}_{\mathcal{I}_1}(t)$ 
since the denominator is just a normalizing constant. $\forall t \in \mathcal{T}$, the sequence $\{\max_{j \in \mathcal{H}_1} |y_{j}(t)- g_{\mathcal{I}_1}(t)|\}_{m}$ is eventually bounded by $q_{1-\alpha}$ and is eventually increasing since $\{\vert \mathcal{H}_1 \vert\}_m$ is eventually increasing. By the monotone convergence theorem, the sequence converges to its supremum.

In order to prove the convergence of the numerator of $\bar{s}^{c}_{\mathcal{I}_1}$ to the same limit function, it is sufficient to consider the previous computations by noting that if $n \to +\infty$ then $l=n(1-\theta) \to +\infty$ and by substituting  $\gamma$ with $k$, $m$ with $l$, $\mathcal{H}_1$ with $\mathcal{H}_2$ and $\mathcal{I}_1$ with $\mathcal{I}_2$ (except for $g_{\mathcal{I}_1}$ that is naturally not substituted by $g_{\mathcal{I}_2}$). Since the numerators of $\bar{s}_{\mathcal{I}_1}$ and $\bar{s}^{c}_{\mathcal{I}_1}$ converge to the same function, also the two normalizing constants converge to the same quantity. In view of this and since $\mathcal{C}^{\bar{s}}_{n, 1-\alpha}$ and $ \mathcal{C}^{\bar{s}^c}_{n, 1-\alpha}$ are defined as
\begin{align*}
\mathcal{C}^{\bar{s}}_{n, 1-\alpha}&:= \left\{y \in \mathcal{Y(T)}: y(t) \in [g_{\mathcal{I}_1}(t)-k^{\bar{s}} \bar{s}_{\mathcal{I}_1}(t) , g_{\mathcal{I}_1}(t)+k^{\bar{s}} \bar{s}_{\mathcal{I}_1}(t)] \quad \forall t \in \mathcal{T}  \right\}, \\
\mathcal{C}^{\bar{s}^c}_{n, 1-\alpha}&:= \left\{y \in \mathcal{Y(T)}: y(t) \in [g_{\mathcal{I}_1}(t)-k^{\bar{s}^c} \bar{s}^c_{\mathcal{I}_1}(t) , g_{\mathcal{I}_1}(t)+k^{\bar{s}^c} \bar{s}^c_{\mathcal{I}_1}(t)] \quad \forall t \in \mathcal{T}  \right\}
\end{align*}
 then $\lim_{n \to +\infty} \mathcal{C}^{\bar{s}}_{n, 1-\alpha}= \lim_{n \to +\infty} \mathcal{C}^{\bar{s}^c}_{n, 1-\alpha}$.

\textbf{Proof of Theorem \ref{th:better_than_not_modul}}. 

The proof consists of two steps. At the first step we show that $k^{\bar{s}^{c}}=\int_{\mathcal{T}} \max_{j \in \mathcal{H}_2} |y_{j}(t)- g_{\mathcal{I}_1}(t)| dt$, a fundamental result to obtain, at the second step, the proof of the theorem.

\textit{I step}

In order not to overcomplicate the proof, first of all let us consider the case in which $\vert \mathcal{H}_2 \vert = \lceil (l+1)(1-\alpha) \rceil $. It is important to notice that under the assumption concerning the continuous joint distribution of $\{ R_h : h \in \mathcal{I}_2\}$ made in Section \ref{sec:conformal_approach}  such condition is always satisfied. However, the result proved at this first step holds also when this assumption is violated, and its proof requires just minor changes. Therefore, for the sake of completeness such proof  is addressed below.

\begin{itemize}
\item $\forall i \in \mathcal{H}_2$ the following relationship holds $\forall t \in \mathcal{T}$:
\begin{align*}
\phantom{=}& \left| \frac{y_{i}(t)- g_{\mathcal{I}_1}(t)}{\bar{s}^{c}_{\mathcal{I}_1}(t)} \right| \\
=& \int_{\mathcal{T}} \max_{j \in \mathcal{H}_2} |y_{j}(t)- g_{\mathcal{I}_1}(t)| dt  \cdot   \frac{\left| y_{i}(t)- g_{\mathcal{I}_1}(t) \right|}{\max_{j \in \mathcal{H}_2} |y_{j}(t)- g_{\mathcal{I}_1}(t)|}  \\
\leq& \int_{\mathcal{T}} \max_{j \in \mathcal{H}_2} |y_{j}(t)- g_{\mathcal{I}_1}(t)| dt,
\end{align*}
and then
\begin{align*}
R^{\bar{s}^{c}}_i:= \sup_{t \in \mathcal{T}} \left| \frac{y_{i}(t)- g_{\mathcal{I}_1}(t)}{\bar{s}^{c}_{\mathcal{I}_1}(t)} \right| \leq \int_{\mathcal{T}} \max_{j \in \mathcal{H}_2} |y_{j}(t)- g_{\mathcal{I}_1}(t)| dt.
\end{align*}

Specifically, $\exists$ $\underline{i} \in  \mathcal{H}_2$ such that $R^{\bar{s}^{c}}_{\underline{i}}= \int_{\mathcal{T}} \max_{j \in \mathcal{H}_2} |y_{j}(t)- g_{\mathcal{I}_1}(t)| dt $ since $\forall t \in \mathcal{T}$ at least one function $y_{\underline{i}}$ satisfies $|y_{\underline{i}}(t)- g_{\mathcal{I}_1}(t)|=\max_{j \in \mathcal{H}_2} |y_{j}(t)- g_{\mathcal{I}_1}(t)|$.

\item Let us define  $\mathcal{CH}_2:=\mathcal{I}_2 \setminus \mathcal{H}_2$  and let $t^{*}_i$ be the value such that 
\begin{equation*} 
|y_{i}(t^{*}_i)- g_{\mathcal{I}_1}(t^{*}_i)|= \sup_{t \in \mathcal{T}} |y_{i}(t)- g_{\mathcal{I}_1}(t)| \quad \forall i \in \mathcal{I}_2.
\end{equation*} 
If $t^*_i$ is not unique, it is randomly chosen from the values that satisfy that condition. $\forall i \in \mathcal{CH}_2$,  by definition of $\mathcal{H}_2$ we obtain that $\left| y_{i}(t^{*}_i)- g_{\mathcal{I}_1}(t^{*}_i) \right| > \max_{j \in \mathcal{H}_2} |y_{j}(t^{*}_i)- g_{\mathcal{I}_1}(t^{*}_i)|$ and so  the following relationship holds:
\begin{align*}
\phantom{=}& \left| \frac{y_{i}(t^{*}_i)- g_{\mathcal{I}_1}(t^{*}_i)}{\bar{s}^{c}_{\mathcal{I}_1}(t^{*}_i)} \right| \\
=& \int_{\mathcal{T}} \max_{j \in \mathcal{H}_2} |y_{j}(t)- g_{\mathcal{I}_1}(t)| dt  \cdot   \frac{\left| y_{i}(t^{*}_i)- g_{\mathcal{I}_1}(t^{*}_i) \right|}{\max_{j \in \mathcal{H}_2} |y_{j}(t^{*}_i)- g_{\mathcal{I}_1}(t^{*}_i)|}  \\
>& \int_{\mathcal{T}} \max_{j \in \mathcal{H}_2} |y_{j}(t)- g_{\mathcal{I}_1}(t)| dt. 
\end{align*}
As a consequence, 
\begin{align*}
R^{\bar{s}^{c}}_i:= \sup_{t \in \mathcal{T}} \left| \frac{y_{i}(t)- g_{\mathcal{I}_1}(t)}{\bar{s}^{c}_{\mathcal{I}_1}(t)} \right| > \int_{\mathcal{T}} \max_{j \in \mathcal{H}_2} |y_{j}(t)- g_{\mathcal{I}_1}(t)| dt.
\end{align*}
\end{itemize}

Since:
\begin{itemize}
\item $\vert \mathcal{H}_2 \vert = \lceil (l+1)(1-\alpha) \rceil $
\item $\forall i \in \mathcal{H}_2$ $R^{\bar{s}^{c}}_i \leq \int_{\mathcal{T}} \max_{j \in \mathcal{H}_2} |y_{j}(t)- g_{\mathcal{I}_1}(t)| dt$ and  $\exists$ $\underline{i} \in  \mathcal{H}_2$ such that $R^{\bar{s}^{c}}_{\underline{i}}= \int_{\mathcal{T}} \max_{j \in \mathcal{H}_2} |y_{j}(t)- g_{\mathcal{I}_1}(t)| dt $
\item $\forall i \in \mathcal{CH}_2$ $R^{\bar{s}^{c}}_i > \int_{\mathcal{T}} \max_{j \in \mathcal{H}_2} |y_{j}(t)- g_{\mathcal{I}_1}(t)| dt$ 
\end{itemize}

we conclude that $k^{\bar{s}^{c}}=\int_{\mathcal{T}} \max_{j \in \mathcal{H}_2} |y_{j}(t)- g_{\mathcal{I}_1}(t)| dt$, with $k^{\bar{s}^{c}}$ the $\lceil (l+1)(1-\alpha) \rceil$th smallest value in the set $\{R^{\bar{s}^{c}}_h: h \in \mathcal{I}_2 \}$.

If $\vert \mathcal{H}_2 \vert > \lceil (l+1)(1-\alpha) \rceil $, then $R^{\bar{s}^{c}}_{i}= \int_{\mathcal{T}} \max_{j \in \mathcal{H}_2} |y_{j}(t)- g_{\mathcal{I}_1}(t)| dt $ is valid $\forall i \in \mathcal{H}_2$ such that $\sup_{t \in \mathcal{T}} \left|y_{i}(t)- g_{\mathcal{I}_1}(t)\right|=k$ 
and in the same way we can conclude that $k^{\bar{s}^{c}}= \int_{\mathcal{T}} \max_{j \in \mathcal{H}_2} |y_{j}(t)- g_{\mathcal{I}_1}(t)| dt$.

\textit{II step}

Let us define $\forall i \in \mathcal{I}_2$
\begin{equation*}
R^{s^{0}}_i:= \sup_{t \in \mathcal{T}} \left| \frac{y_{i}(t)- g_{\mathcal{I}_1}(t)}{s^{0}(t)} \right| =  \left| \mathcal{T} \right|  \sup_{t \in \mathcal{T}} \left| y_{i}(t)- g_{\mathcal{I}_1}(t) \right|.
\end{equation*}
Since $k^{s^{0}}$ is the $\lceil (l+1)(1-\alpha) \rceil$th smallest value in the set $\{R^{s^{0}}_h: h \in \mathcal{I}_2 \}$, by definition of $\mathcal{H}_2$ we obtain that
\begin{align*}
k^{s^{0}} =&   \left| \mathcal{T} \right| \max_{j \in \mathcal{H}_2} \left( \sup_{t \in \mathcal{T}} \left| y_{j}(t)- g_{\mathcal{I}_1}(t) \right| \right)\\
=&   \left| \mathcal{T} \right| \sup_{t \in \mathcal{T}} \left( \max_{j \in \mathcal{H}_2}  \left| y_{j}(t)- g_{\mathcal{I}_1}(t) \right| \right).
\end{align*}

Since at the first step we proved that $k^{\bar{s}^{c}}=\int_{\mathcal{T}} \max_{j \in \mathcal{H}_2} |y_{j}(t)- g_{\mathcal{I}_1}(t)| dt$, we obtain that
\begin{equation*}
k^{s^{0}} - k^{\bar{s}^{c}} = \left| \mathcal{T} \right| \sup_{t \in \mathcal{T}} \left( \max_{j \in \mathcal{H}_2}  \left| y_{j}(t)- g_{\mathcal{I}_1}(t) \right| \right) - \int_{\mathcal{T}} \max_{j \in \mathcal{H}_2} |y_{j}(t)- g_{\mathcal{I}_1}(t)| dt.
\end{equation*}

Since the right side of the equation is greater than or equal to 0 by the integral mean value theorem, then $\mathcal{Q}(s^{0}) \geq \mathcal{Q}( \bar{s}^{c}_{\mathcal{I}_1})$.

The same theorem ensures that 
\begin{align*}
 \left| \mathcal{T} \right|  \sup_{t \in \mathcal{T}} \left( \max_{j \in \mathcal{H}_2}  \left| y_{j}(t)- g_{\mathcal{I}_1}(t) \right| \right) = \int_{\mathcal{T}} \max_{j \in \mathcal{H}_2} |y_{j}(t)- g_{\mathcal{I}_1}(t)| dt \\
\iff  \max_{j \in \mathcal{H}_2} |y_{j}(t)- g_{\mathcal{I}_1}(t)| \quad \text{is constant almost everywhere},
\end{align*}

i.e. if and only if $  \bar{s}^{c}_{\mathcal{I}_1}(t) = \bar{s}^{0}(t)$ almost everywhere.

\textbf{Proof of Theorem \ref{th:better_than_group_of_functions}}.

We have already shown at the first step of the previous proof that $k^{\bar{s}^{c}}=\int_{\mathcal{T}} \max_{j \in \mathcal{H}_2} |y_{j}(t)- g_{\mathcal{I}_1}(t)| dt$.
Since by assumption $s^d_{\mathcal{I}_1}(t^{*}_i) \leq \bar{s}^c_{\mathcal{I}_1}(t^{*}_i) $ $\forall i \in \mathcal{CH}_2$ and $\vert \mathcal{H}_2 \vert = \lceil (l+1)(1-\alpha) \rceil$, let us define $a_i \geq 0$  $\forall i \in \mathcal{CH}_2$ the value such that $s^d_{\mathcal{I}_1}(t^{*}_i)= \bar{s}^c_{\mathcal{I}_1}(t^{*}_i)-a_i$.

\begin{itemize}
\item \textit{Case 1}: If  $\exists$ $x \in \mathcal{CH}_2$ s.t. $a_{x} > 0 $, $\exists$ $\underline{i} \in  \mathcal{H}_2$ such that
\begin{align*}
\phantom{=}& \left| \frac{y_{{\underline{i}}}(t^{*}_{x})- g_{\mathcal{I}_1}(t^{*}_{x})}{s^{d}_{\mathcal{I}_1}(t^{*}_{x})} \right| \\
=& \left| \frac{y_{{\underline{i}}}(t^{*}_{x})- g_{\mathcal{I}_1}(t^{*}_x)}{ \bar{s}^c_{\mathcal{I}_1}(t^{*}_{x})-a_{x}} \right| \\
=& \int_{\mathcal{T}} \max_{j \in \mathcal{H}_2} |y_{j}(t)- g_{\mathcal{I}_1}(t)| dt \quad \times   \\
&\frac{\left| y_{{\underline{i}}}(t^{*}_{x})- g_{\mathcal{I}_1}(t^{*}_{x}) \right|}{\max_{j \in \mathcal{H}_2} |y_{j}(t^{*}_{x})- g_{\mathcal{I}_1}(t^{*}_{x})| -a_{x} \cdot \int_{\mathcal{T}} \max_{j \in \mathcal{H}_2} |y_{j}(t)- g_{\mathcal{I}_1}(t)| dt}  \\
>& \int_{\mathcal{T}} \max_{j \in \mathcal{H}_2} |y_{j}(t)- g_{\mathcal{I}_1}(t)| dt
\end{align*}

since $\forall t \in \mathcal{T}$ (and specifically for $t^{*}_{x}$) at least one function $y_{\underline{i}}$ satisfies $|y_{\underline{i}}(t)- g_{\mathcal{I}_1}(t)|=\max_{j \in \mathcal{H}_2} |y_{j}(t)- g_{\mathcal{I}_1}(t)|$. 

\textit{Case 2}: If $a_i = 0$ $\forall i \in \mathcal{CH}_2$, there exist at least two values $t_{\downarrow},t_{\uparrow} \in \mathcal{T}^*$ such that $s^d_{\mathcal{I}_1}(t_{\downarrow}) < \bar{s}^c_{\mathcal{I}_1}(t_{\downarrow}) $ and $s^d_{\mathcal{I}_1}(t_{\uparrow}) > \bar{s}^c_{\mathcal{I}_1}(t_{\uparrow}) $ since otherwise $s^d_{\mathcal{I}_1}(t) = \bar{s}^c_{\mathcal{I}_1}(t)$ $\forall t \in \mathcal{T}^*$. Let us define $a_{\downarrow}>0$ the value such that $s^d_{\mathcal{I}_1}(t_{\downarrow}) = \bar{s}^c_{\mathcal{I}_1}(t_{\downarrow})-a_{\downarrow} $.
Therefore $\exists$ $\underline{i} \in  \mathcal{H}_2$ such that
\begin{align*}
\phantom{=}& \left| \frac{y_{{\underline{i}}}(t_{\downarrow})- g_{\mathcal{I}_1}(t_{\downarrow})}{s^{d}_{\mathcal{I}_1}(t_{\downarrow})} \right| \\
=& \left| \frac{y_{{\underline{i}}}(t_{\downarrow})- g_{\mathcal{I}_1}(t_{\downarrow})}{ \bar{s}^c_{\mathcal{I}_1}(t_{\downarrow})-a_{\downarrow}} \right| \\
=& \int_{\mathcal{T}} \max_{j \in \mathcal{H}_2} |y_{j}(t)- g_{\mathcal{I}_1}(t)| dt \quad \times   \\
&\frac{\left| y_{{\underline{i}}}(t_{\downarrow})- g_{\mathcal{I}_1}(t_{\downarrow}) \right|}{\max_{j \in \mathcal{H}_2} |y_{j}(t_{\downarrow})- g_{\mathcal{I}_1}(t_{\downarrow})| -a_{\downarrow} \cdot \int_{\mathcal{T}} \max_{j \in \mathcal{H}_2} |y_{j}(t)- g_{\mathcal{I}_1}(t)| dt}  \\
>& \int_{\mathcal{T}} \max_{j \in \mathcal{H}_2} |y_{j}(t)- g_{\mathcal{I}_1}(t)| dt
\end{align*}
since $\forall t \in \mathcal{T}$ (and specifically for $t_{\downarrow}$) at least one function $y_{\underline{i}}$ satisfies $|y_{\underline{i}}(t)- g_{\mathcal{I}_1}(t)|=\max_{j \in \mathcal{H}_2} |y_{j}(t)- g_{\mathcal{I}_1}(t)|$. 

As a consequence, in both cases ($\exists x \in \mathcal{CH}_2$ s.t. $a_x > 0 $ and $a_i = 0$ $\forall i \in \mathcal{CH}_2$) we obtain that $\exists$ $\underline{i} \in  \mathcal{H}_2$ such that
\begin{align*}
R^{s^{d}}_{\underline{i}}:= \sup_{t \in \mathcal{T}} \left| \frac{y_{{\underline{i}}}(t)- g_{\mathcal{I}_1}(t)}{s^{d}_{\mathcal{I}_1}(t)} \right| > \int_{\mathcal{T}} \max_{j \in \mathcal{H}_2} |y_{j}(t)- g_{\mathcal{I}_1}(t)| dt.
\end{align*}

\item $\forall i \in \mathcal{CH}_2$,  by definition of $\mathcal{H}_2$ we obtain that $\left| y_{i}(t^{*}_i)- g_{\mathcal{I}_1}(t^{*}_i) \right| > \max_{j \in \mathcal{H}_2} |y_{j}(t^{*}_i)- g_{\mathcal{I}_1}(t^{*}_i)|$ and so  the following relationship holds:
\begin{align*}
\phantom{=}& \left| \frac{y_{i}(t^{*}_i)- g_{\mathcal{I}_1}(t^{*}_i)}{s^{d}_{\mathcal{I}_1}(t^{*}_i)} \right| \\
=& \left| \frac{y_{i}(t^{*}_i)- g_{\mathcal{I}_1}(t^{*}_i)}{ \bar{s}^c_{\mathcal{I}_1}(t^{*}_i)-a_i} \right| \\
=& \int_{\mathcal{T}} \max_{j \in \mathcal{H}_2} |y_{j}(t)- g_{\mathcal{I}_1}(t)| dt \quad \times   \\
&\frac{\left| y_{i}(t^{*}_i)- g_{\mathcal{I}_1}(t^{*}_i) \right|}{\max_{j \in \mathcal{H}_2} |y_{j}(t^{*}_i)- g_{\mathcal{I}_1}(t^{*}_i)| -a_j \cdot \int_{\mathcal{T}} \max_{j \in \mathcal{H}_2} |y_{j}(t)- g_{\mathcal{I}_1}(t)| dt}  \\
>& \int_{\mathcal{T}} \max_{j \in \mathcal{H}_2} |y_{j}(t)- g_{\mathcal{I}_1}(t)| dt.
\end{align*}
As a consequence, 
\begin{align*}
R^{s^{d}}_i:= \sup_{t \in \mathcal{T}} \left| \frac{y_{i}(t)- g_{\mathcal{I}_1}(t)}{s^{d}_{\mathcal{I}_1}(t)} \right| > \int_{\mathcal{T}} \max_{j \in \mathcal{H}_2} |y_{j}(t)- g_{\mathcal{I}_1}(t)| dt.
\end{align*}
\end{itemize}
Since:
\begin{itemize}
\item $\vert \mathcal{H}_2 \vert = \lceil (l+1)(1-\alpha) \rceil $
\item  $\exists$ $\underline{i} \in  \mathcal{H}_2$ such that $R^{s^{d}}_{\underline{i}} > \int_{\mathcal{T}} \max_{j \in \mathcal{H}_2} |y_{j}(t)- g_{\mathcal{I}_1}(t)| dt $
\item $\forall i \in \mathcal{CH}_2$ $R^{s^{d}}_i > \int_{\mathcal{T}} \max_{j \in \mathcal{H}_2} |y_{j}(t)- g_{\mathcal{I}_1}(t)| dt$
\end{itemize}

we conclude that $k^{s^{d}} > \int_{\mathcal{T}} \max_{j \in \mathcal{H}_2} |y_{j}(t)- g_{\mathcal{I}_1}(t)| dt$, i.e. $k^{s^{d}} >  k^{\bar{s}^{c}}$,  with $k^{s^{d}}$ the $\lceil (l+1)(1-\alpha) \rceil$th smallest value in the set $\{R^{s^{d}}_h: h \in \mathcal{I}_2 \}$.
.

\textbf{Proof that Theorem \ref{th:better_than_group_of_functions} does not imply Theorem \ref{th:better_than_not_modul}}. 

Theorem \ref{th:better_than_group_of_functions} does not imply Theorem \ref{th:better_than_not_modul} since $s^{0}$ may not fulfill $s^0(t^{*}_i) \leq \bar{s}^c_{\mathcal{I}_1}(t^{*}_i) $ $\forall i \in \mathcal{CH}_2$. In fact, $\forall i \in \mathcal{CH}_2$:
\begin{equation*}
s^0(t^{*}_i) \leq \bar{s}^c_{\mathcal{I}_1}(t^{*}_i)  \iff  \frac{\int_{\mathcal{T}} \max_{j \in \mathcal{H}_2} |y_{j}(t)- g_{\mathcal{I}_1}(t)| dt}{\left| \mathcal{T} \right|} \leq \max_{j \in \mathcal{H}_2}  \left| y_{j}(t^{*}_i)- g_{\mathcal{I}_1}(t^{*}_i) \right|
\end{equation*}
and the condition on the right side is not always satisfied because no constraints are imposed on $y_{j}(t_i^*)$, with  $j \in \mathcal{H}_2$, $i \in \mathcal{CH}_2$.

\subsection*{A.4 Proofs about Smoothed Conformal Predictor}

\textbf{Proof of the smoothed conformal prediction set }

By considering the notation of Section \ref{sec:conformal_prediction_bands}, first of all let us notice that, by definition, $\mathcal{C}_{n, 1-\alpha,1}= \mathcal{C}_{n, 1-\alpha}$.

Since $\delta_{y,\tau_{n+1}}$ can not be less than $\tau_{n+1}/(l+1)$ and can not be greater than $(l+\tau_{n+1})/(l+1)$, we consider the case in which $\alpha \in [\tau_{n+1}/(l+1), (l+\tau_{n+1})/(l+1) )$. Let us define $w$ the $\lceil l + \tau_{n+1} - (l+1) \alpha  \rceil$th smallest value in the set $\{R_h: h \in \mathcal{I}_2 \}$, and $r_{n}$ ($v_n$ respectively) the number of elements in the set $\{R_h: h \in \mathcal{I}_2 \}$ that are equal to $w$ and that are to the right (left respectively) of $w$ in the sorted version of the set.
Under the assumption concerning the continuous joint distribution of $\{ R_h : h \in \mathcal{I}_2\}$ made in Section \ref{sec:conformal_approach}  $r_{n}=v_{n}=0$ holds, but generally speaking we assume $r_{n},v_{n} \in \mathcal{N}_{\geq0}$ such that $r_{n}+v_{n} \leq l-1$. By performing calculations similar to those needed in the non-randomized scenario, we obtain that:
\begin{itemize}
\item if 
\begin{equation*}
\tau_{n+1}>\frac{(l+1)\alpha-\lfloor (l+1) \alpha -\tau_{n+1} \rfloor + r_{n}}{r_{n} + v_{n} + 2}
\end{equation*}
then $y \in \mathcal{C}_{n, 1-\alpha,\tau_{n+1}} \iff R_{n+1} \leq w$ and so
\begin{align*}
\mathcal{C}_{n, 1-\alpha,\tau_{n+1}} = \{ y \in \mathcal{Y(T)}: y(t) \in [&g_{\mathcal{I}_1}(t)- w  ,  \\
                                                                                                             &g_{\mathcal{I}_1}(t)+w] \quad \forall t \in \mathcal{T}  \}
\end{align*}

\item if 
\begin{equation*}
\tau_{n+1} \leq \frac{(l+1)\alpha-\lfloor (l+1) \alpha -\tau_{n+1} \rfloor + r_{n}}{r_{n} + v_{n} + 2}
\end{equation*}
then $y \in \mathcal{C}_{n, 1-\alpha,\tau_{n+1}} \iff R_{n+1} < w$ and so
\begin{align*}
\mathcal{C}_{n, 1-\alpha,\tau_{n+1}} = \{ y \in \mathcal{Y(T)}: y(t) \in \bigl(&g_{\mathcal{I}_1}(t)- w ,  \\
                                                                                                             &g_{\mathcal{I}_1}(t)+w\bigl) \quad \forall t \in \mathcal{T}  \}.
\end{align*}
\end{itemize}

Also the introduction of the modulation function presented in Section \ref{sec:modulation} can be easily generalized in the smoothed conformal context. Let us define for a given $y \in \mathcal{Y(T)}$
\begin{align*}
\delta^s_{y,\tau_{n+1}} &:=  \frac{\left|\left\{j \in  \mathcal{I}_2: R^s_{j} > R^s_{n+1}\right\}\right| + \tau_{n+1} \left|\left\{j \in  \mathcal{I}_2 \cup \{n+1\}: R^s_{j} = R^s_{n+1}\right\}\right|}{l+1} \\
\mathcal{C}_{n, 1-\alpha,\tau_{n+1}}^{s}&:= \left\{y \in \mathcal{Y(T)}: \delta^{s}_{y, \tau_{n+1}}>\alpha \right\}.
\end{align*}
By reconsidering the previous computations and by substituting $\delta_{y,\tau_{n+1}}$ with $\delta^s_{y,\tau_{n+1}}$, $w$ with $w^s$, $R_h$ with $R_h^s$, $r_{n}$ with $r_{n}^s$ and $v_{n}$ with $v^s_{n}$ it is possible to notice that 

\begin{itemize}
\item if 
\begin{equation*}
\tau_{n+1}>\frac{(l+1)\alpha-\lfloor (l+1) \alpha -\tau_{n+1} \rfloor + r^s_{n}}{r^s_{n} + v^s_{n} + 2}
\end{equation*}
then 
\begin{align*}
\mathcal{C}_{n, 1-\alpha,\tau_{n+1}}^{s} = \{ y \in \mathcal{Y(T)}: y(t) \in [&g_{\mathcal{I}_1}(t)- w^s s_{\mathcal{I}_1}(t) ,  \\
                                                                                                             &g_{\mathcal{I}_1}(t)+w^s s_{\mathcal{I}_1}(t)] \quad \forall t \in \mathcal{T}  \}
\end{align*}

\item if 
\begin{equation*}
\tau_{n+1} \leq \frac{(l+1)\alpha-\lfloor (l+1) \alpha -\tau_{n+1} \rfloor + r^s_{n}}{r^s_{n} + v^s_{n} + 2}
\end{equation*}
then 
\begin{align*}
\mathcal{C}_{n, 1-\alpha,\tau_{n+1}}^{s} = \{ y \in \mathcal{Y(T)}: y(t) \in \bigl(&g_{\mathcal{I}_1}(t)- w^s s_{\mathcal{I}_1}(t) ,  \\
                                                                                                             &g_{\mathcal{I}_1}(t)+w^s s_{\mathcal{I}_1}(t)\bigl) \quad \forall t \in \mathcal{T}  \}.
\end{align*}
\end{itemize}

\textbf{Proof of Remark \ref{remark:smoothed1}}. 

The functions $ \bar{s}^{c}_{\mathcal{I}_1}$ and $ \bar{s}_{\mathcal{I}_1}$ are defined as in Section \ref{sec:modulation} except for $k$ ($\gamma$ respectively) that is the $\lceil l + \tau_{n+1} - (l+1) \alpha  \rceil$th ($\lceil m + \tau_{n+1} - (m+1) \alpha  \rceil$th respectively) smallest value in the corresponding set; similarly, if $\lceil m + \tau_{n+1} - (m+1) \alpha \rceil > m$ then $\mathcal{H}_1=\mathcal{I}_1$ and if $\lceil m + \tau_{n+1} - (m+1) \alpha \rceil \leq 0$ we arbitrarily set $\bar{s}_{\mathcal{I}_1}=s^0$. The theorems of Section  \ref{sec:modulation} still hold by substituting $\lceil (l+1)(1-\alpha) \rceil, \lceil (m+1)(1-\alpha) \rceil$ with $\lceil l + \tau_{n+1} - (l+1) \alpha  \rceil, \lceil m + \tau_{n+1} - (m+1) \alpha  \rceil$.

\section*{Acknowledgements}

Prof. Vantini and Dr. Fontana acknowledge the financial support from Accordo Quadro ASI-POLIMI ``Attivit\`a di Ricerca e Innovazione'' n. 2018-5-HH.0, collaboration agreement between the Italian Space Agency and Politecnico di Milano. The authors would like to thank Giulio Pegorer for fruitful discussions.

\bibliographystyle{agsm}

\end{document}